\documentclass[twocolumn, journal]{IEEEtran}
\usepackage{amsmath,amsfonts,amssymb,amsthm,mathtools,amsfonts}
\usepackage{color}
\usepackage{algpseudocode}
\usepackage{algorithm}
\usepackage{array}
\usepackage{mathtools}

\usepackage{stfloats}
\usepackage{url}
\usepackage{verbatim}
\usepackage{graphicx}
\usepackage{cite}
\usepackage{amsmath}
\usepackage{bbm}
\usepackage{cite}
\usepackage{kotex}
\usepackage{subfigure}

\makeatletter
\def\equalsfill{$\m@th\mathord=\mkern-7mu
\cleaders\hbox{$\!\mathord=\!$}\hfill
\mkern-7mu\mathord=$}
\makeatother

\newtheorem{definition}{Definition}
\newtheorem{lemma}{Lemma}
\newtheorem{theorem}{Theorem}
\newtheorem{remark}{Remark}

\begin{document}

\title{Privacy-Enhanced Over-the-Air Federated Learning via Client-Driven Power Balancing}

\author{
	\IEEEauthorblockN{Bumjun Kim},~\IEEEmembership{Graduate Student Member,~IEEE},
	\IEEEauthorblockN{Hyowoon Seo},~\IEEEmembership{Member,~IEEE},
	and
	\IEEEauthorblockN{Wan Choi},~\IEEEmembership{Fellow,~IEEE}
	\vspace{-0.1in}
	\thanks{This work was supported by the National Research Foundation of Korea(NRF) grant funded by the Korea government (MSIT) (RS-2021-NR059011). \emph{(Corresponding authors: Hyowoon Seo and Wan Choi.)}}
        \thanks{B. Kim and W.~Choi are with the Department of Electrical and Computer Engineering and the Institute of New Media and Communications, Seoul National University
		(SNU), Seoul 08826, Korea (e-mail: \{eithank96,~wanchoi\}@snu.ac.kr).}
  		\thanks{Hyowoon Seo is with the Department of Electrical and Computer Engineering, Sungkyunkwan University, Suwon 16419, South Korea (e-mail: hyowoonseo@skku.edu).} 
		
}

\maketitle
\begin{abstract}
This paper introduces a novel privacy-enhanced over-the-air Federated Learning (OTA-FL) framework using client-driven power balancing (CDPB) to address privacy concerns in OTA-FL systems. In recent studies, a server determines the power balancing based on the continuous transmission of channel state information (CSI) from each client. Furthermore, they concentrate on fulfilling privacy requirements in every global iteration, which can heighten the risk of privacy exposure as the learning process extends. To mitigate these risks, we propose two CDPB strategies—CDPB-n (noisy) and CDPB-i (idle)—allowing clients to adjust transmission power independently, without sharing CSI. CDPB-n transmits noise during poor conditions, while CDPB-i pauses transmission until conditions improve. To further enhance privacy and learning efficiency, we show a mixed strategy, CDPB-mixed, which combines CDPB-n and CDPB-i. Our experimental results show that CDPB outperforms traditional approaches in terms of model accuracy and privacy guarantees providing a practical solution for enhancing OTA-FL in resource-constrained environments.
\end{abstract}

\IEEEpeerreviewmaketitle
\section{Introduction}
Federated Learning (FL) represents a paradigm shift from traditional centralized data collection and analysis methodologies, emphasizing data privacy across distributed client networks while enabling effective training of integrated machine learning (ML) models\cite{mcmahan2017communication ,kairouz2021advances, 10038545}. This framework centralizes around an algorithmic process that aggregates neural network (NN) models independently developed by clients through a central parameter server. These models are subsequently merged into a global model and then redistributed to the clients. Typically, the aggregation of local models in FL demands communication resources that scale with the number of clients. However, recent advancements in FL over wireless communication networks exploit the superposition property for averaging during the aggregation process through a multiple access channel (MAC), significantly reducing communication resource requirements\cite{yang2020federated,9743383}. This innovative approach, known as over-the-air FL (OTA-FL), is currently a focal point of extensive research across multiple scientific disciplines, highlighting its potential to transform the field of wireless distributed ML.

Despite its benefits, the FL paradigm remains vulnerable to privacy risks during the exchange of local models, potentially compromising client data confidentiality \cite{kairouz2021advances}. For instance, sophisticated attacks like model inversion and membership inference can exploit local gradients to either reconstruct original input data or verify client-specific data usage, posing significant threats \cite{huang2021evaluating, shokri2017membership}. To mitigate these vulnerabilities, differential privacy (DP) has been introduced as a vital safeguard. This mathematical framework quantifies the privacy an algorithm offers and minimizes intrusions from its outputs. In FL, privacy can be preserved by integrating a controlled amount of artificially generated Gaussian noise into the model updates before dissemination, effectively concealing individual data contributions while maintaining collective learning integrity \cite{Mohamed_amplification, liu2020privacy}. More recently, the R\'enyi Differential Privacy (RDP) framework was introduced, extending the traditional DP model \cite{mironov2017renyi}. By incorporating a subsampling mechanism, which selects a random subset of clients or data points in each training round, the RDP framework tightened privacy bounds by limiting data exposure during each iteration. The integration of RDP is particularly beneficial in FL environments, where datasets are processed through a series of randomized mechanisms. This framework facilitates a more accurate quantification of DP metrics, enhancing the assessment of privacy guarantees within FL systems.

Similarly, recent studies on OTA-FL have employed the DP concept to enhance privacy \cite{liu2020privacy, seif2020wireless, liu2024differentially,deno2023, 10264827,yan2023over,mao2024leveraging,liao2022over,xue2023over}. As a pioneering study, the authors in \cite{seif2020wireless} proposed a private wireless gradient aggregation scheme in OTA-FL. Their approach achieves a privacy leakage per client that scales as $O(1/\sqrt{K})$, providing a significant improvement in privacy protection. In \cite{liu2020privacy}, a dynamic power allocation scheme for FL is proposed using both orthogonal multiple access (OMA) and non-orthogonal multiple access (NOMA) protocols. Their goal is to minimize the learning convergence error under privacy and power constraints across a given number of communication blocks. Building on these foundations, \cite{deno2023} introduced a framework that jointly optimizes client-side transmission power and server-side denoising factors under fading channels, thereby enhancing the trade-off between convergence error and DP. Moreover, \cite{10264827} provides an analysis of DP in OTA‑FL by taking into account the inherent randomness in the local gradients, which offers further insights into the stochastic nature of privacy leakage in these systems.

Additionally, \cite{yan2023over} investigated the joint effects of device scheduling and power allocation between gradients and noise within a sum power constraint, analyzing both convergence error and DP.  \cite{mao2024leveraging} proposed a method that leverages inherent channel noise and, when necessary, introduces artificial noise through a cooperative jammer. Furthermore, works such as \cite{liao2022over} and \cite{xue2023over} considered scenarios involving honest-but-curious adversaries. In these studies, the noise added to the clients is designed to be either spatially correlated or pairwise cancellable. This design ensures that artificial noise is eliminated during aggregation on the server while remaining effective against external eavesdroppers. These studies assume that the parameter server has complete knowledge of the channel information between itself and each client. Based on this information, a signal-noise power balancing is used, where the server determines the optimal signal powers of the gradients and artificial noise for each client per global iteration, and then communicates these details to the clients.

Although the techniques proposed in the aforementioned studies achieve satisfactory learning accuracy and privacy performance, there are still several issues that need to be addressed. First, they require the transmission of orthogonal feedback of current CSI from each client to the parameter server, and the server must continuously transmit power balancing information to each client, even when the channel distribution remains unchanged. This results in communication overhead. Second, most of these techniques focus on meeting DP requirements in each iteration, which increases the risk of privacy exposure as the learning process extends, potentially leading to greater privacy breaches in successive iterations.\footnote{As the number of iterations increases, the same data points may be used repeatedly across multiple iterations. The exposure of multiple iterations can lead to a gradual and progressive leakage of privacy over time\cite{mironov2017renyi}.}
 
To address these challenges, we introduce a novel privacy-enhanced OTA-FL framework, wherein the parameter server, equipped with only the information of channel distribution between itself and the clients, can perform over-the-air averaging of local gradient updates and enhance privacy via a novel strategy termed \emph{client-driven power balancing} (\texttt{CDPB}). The proposed OTA-FL framework classifies clients into two groups in each iteration, i.e., \emph{Clients with Good channels (CwG)} and \emph{Clients with Poor channels (CwP)}, contingent upon the channel condition measured at each client.

On the one hand, CwG autonomously balances the signal-noise power ratio based on power balancing parameters provided by the server. These parameters, derived from the channel distribution by the server, facilitate a privacy-enhanced OTA-FL process with respect to RDP. Unlike conventional schemes, where power balancing parameters are delivered in every iteration, our approach transmits these parameters only when there is a change in the channel distribution. Furthermore, since the server does not have access to each client's CSI in every iteration, it cannot determine which clients are participating in a given iteration. This uncertainty regarding client participation enhances the overall RDP guarantees, as it introduces an additional layer of privacy protection by obscuring client involvement in each iteration.

On the other hand, two strategies for CwP are considered. The first strategy, termed \texttt{noisy}, involves CwP transmitting an artificial noise signal solely at maximum power to obscure their transmissions, thus enhancing the network's overall privacy. The second strategy, \texttt{idle}, entails CwP pausing their participation during poor channel conditions and resuming transmissions only when the channel gain reaches an acceptable level. For each strategy, we derive analytical results for convergence error and RDP. Based on these results, we formulate an optimization problem for \texttt{CDPB} with each strategy and determine the optimal power balancing method. Additionally, we calculate the minimum number of training iterations required to achieve convergence in the proposed method, optimizing the balance between learning efficiency and the privacy risk associated with extended training durations.

To validate the efficacy of the proposed privacy-enhanced OTA-FL framework, we conduct experiments comparing baseline methodologies with our proposed approaches: \texttt{CDPB} with \texttt{noisy} CwP (\texttt{CDPB-n}) and \texttt{CDPB} with \texttt{idle} CwP (\texttt{CDPB-i}). The empirical results demonstrate that our methods outperform existing strategies in terms of model accuracy, RDP guarantees, and power efficiency. Additionally, as a secondary objective, we explore the optimal strategy for clients with poor channel conditions, evaluating whether \texttt{CDPB-n} or \texttt{CDPB-i} offers greater advantages. Finally, we introduce and discuss a mixed strategy, \texttt{CDPB-mixed}, which combines both \texttt{CDPB-n} and \texttt{CDPB-i}. By adjusting the proportion of CwP strategies, we show that \texttt{CDPB-mixed} has the potential to enhance the robustness and effectiveness of the OTA-FL system.

\textbf{Contributions.} \quad The key contributions of this article are summarized as follows:
\begin{itemize}
    \item We introduce the \texttt{CDPB} strategy for privacy-enhanced OTA-FL. This method enables clients to independently adjust power between gradient signals and artificial noise based on parameters offered from the server, calculated at the server using channel distribution information. This strategy optimally balances power and minimizes the necessary OTA-FL iterations for convergence.

    \item We examine two privacy strategies for CwP: \texttt{CDPB-n} and \texttt{CDPB-i}. \texttt{CDPB-n} involves CwP transmitting noise at maximum power under low channel gain, while \texttt{CDPB-i} involves pausing transmission. 

    \item We have confirmed the convergence of OTA-FL using \texttt{CDPB-n} and \texttt{CDPB-i}, providing insights into effective power balancing strategies and highlighting the importance of convergence proof to reduce the number of necessary OTA-FL iterations.
    \item Through experimental validation, we demonstrate that the proposed \texttt{CDPB} approach consistently outperforms existing methods. To further leverage the strengths of both strategies, we introduce a mixed approach, \texttt{CDPB-mixed}, which combines their advantages. By adjusting the proportion of CwP transmitting noise, \texttt{CDPB-mixed} improves privacy protections while maintaining efficient convergence.
\end{itemize}

\textbf{Organization.} \quad The remainder of this article is organized as follows. In Section \ref{sec:preliminaries}, we provide a comprehensive review of conventional OTA-FL and RDP. Section \ref{sec:what clients with poor channels} introduces the \texttt{noisy} and \texttt{idle} strategies for CwP with poor channel conditions. In Section \ref{sec:client driven power balancing}, we propose the \texttt{CDPB} strategy to address the optimization problem for privacy-enhanced OTA-FL. Section \ref{sec:experimental results} presents experimental results evaluating the performance of \texttt{CDPB} in comparison with other OTA-FL strategies. Finally, in Section \ref{sec:conclusion}, we conclude and summarize the findings of the article.

\section{Preliminaries}\label{sec:preliminaries}
\subsection{Vanilla Federated Learning Framework}
Consider a FL network composed of a central parameter server and $K$ local clients, with the group of local clients represented by $\mathcal{K}$ and its size given by $|\mathcal{K}| = K$. This network aims to collaboratively develop a model $\theta\in\mathbb{R}^d$ by utilizing the private datasets held by the clients. To achieve a global model $\theta^*$, clients periodically send updates to the server, assuming no direct interaction among the clients. The optimal model $\theta^*$ is identified through the following minimization process:
\begin{align}
    \theta^*=\underset{\theta}{\arg\min}\ f(\theta),
\end{align}
where $f(\theta)=\frac{1}{K}\sum_{\forall k\in\mathcal{K}}f_k(\theta)$ denotes the combined global loss function, and $f_k(\theta)$ indicates the individual local loss function of client $k$, reliant on the dataset $\mathcal{D}_k$ in possession of client $k$. Within the FL framework, the reduction of $f(\theta)$ occurs across several iterations of local and global computations. Specifically, in the Federated Averaging (FedAvg) algorithm \cite{mcmahan2017communication}, the following steps are executed in the $t$-th iteration for $t = 1,2,\dots,\tau$:
\begin{enumerate}
    \item The server distributes the $t$-th iteration global model $\theta_{t}$ to clients by broadcasting.
    \item Each client $k \in \mathcal{K}$ performs $L$ iterations of local stochastic gradient descent (SGD), with $\theta_{k,t,\ell}=\theta_{k,t,\ell-1}-\eta_t\nabla{f_k(\theta_{k,t,\ell})}$ for $\ell = 1,...,L$, resulting in an updated local model $\theta_{k,t,L}$, where $\theta_{k,t,0} = \theta_{t}$.
    \item Client $k \in \mathcal{K}$ returns the updated local gradient $g_{k,t} = \theta_{k,t,L} - \theta_{k,t,0}$ to the server.
    \item The server aggregates and averages the updates to form the $(t+1)$-th iteration global model based on the local modifications from clients.
\end{enumerate}
After $\tau$ iterations of this process, the training concludes, yielding an optimally trained global model. 

\subsection{Over-the-Air Federated Learning (OTA-FL)}\label{sec:OTA-FL}
Now consider a scenario where both the server and clients are engaged in wireless communication for local model aggregation and global model dissemination. In this setup, the aggregation and dissemination processes occur over distinct communication channels: multiple access channel (MAC) for aggregation and broadcast channel (BC) for dissemination. Notably, the throughput for uplink communications over the MAC, which involves data transmission from clients to the server, is typically more constrained than that of downlink communications over the BC. Consequently, our primary focus centers on optimizing uplink communications. Meanwhile, we make the standard assumption in FL studies that reliable communication at arbitrary rates is feasible for the downlink channel \cite{9272666}.

To describe the wireless channels within this network, during the $t$-th iteration of the FL process,  client $k$ experiences a Rayleigh block fading channel denoted as $\sqrt{h_{k,t}} e^{j\phi_{k,t}}$,\footnote{In this work, we eliminate path‐loss bias by assuming standard open‐loop power control. Each client inverts its large-scale fading $\beta_k$ by scaling its transmit amplitude by $\frac{1}{\beta_k}$, resulting in an effective power budget $P_k=\frac{\bar{P}}{\beta_k}$, where $\bar{P}$ is the total power of the clients. Because we consider the variance of $\beta_k$ is small, we set a uniform power budget, i.e., $P_k=P, \forall k.$ In very large or highly heterogeneous networks, clients whose $P_k$ lie in similar ranges can be grouped into clusters and each cluster optimizes the power balancing parameter at its own budget $P^c$, where $c$ is the cluster index.} which follows a complex Gaussian distribution with zero mean and variance $\sigma^2$. Here, $\sqrt{h_{k,t}}$ represents the magnitude, and $e^{j\phi_{k,t}}$ signifies the phase of the channel coefficient between  client $k$ and the server. Let $x_{k,t}$ and $y_t$ respectively represent the transmitted $d \times 1$ vector signal from client $k$ and the received signal at the server. Consequently, we can express the input-output relationship for the MAC channel in the considered FL network during the $t$-th iteration as
\begin{align}\label{eq:channel_model}
y_t = \sum_{k\in\mathcal{K}} \sqrt{h_{k,t}}e^{j\phi_{k,t}}x_{k,t} + z_t,
\end{align}
where, $z_t$ follows a complex Gaussian distribution with zero mean and variance $\sigma_z^2$, representing additive white Gaussian noise. Each term within the summation corresponds to different client $k$ in the set of clients $\mathcal{K}$. Additionally, it is assumed that each client possesses knowledge of both CSI and channel distributions. This information can be acquired from the server through uplink channel estimation. It is also essential to note that each client operates under power constraints, and as such, must adhere to an average power constraint as expressed by the equation
\begin{equation}
\mathbf{E}\left[\lVert x_{k,t}\rVert^2\right] \leq P, \quad \forall(k,t) \in \mathcal{K} \times \mathcal{T},
\end{equation}
where $P>0$ is a positive constant denoting the maximum transmission power available to each client.

One traditional way to aggregate and average local models in the considered wireless environment is to utilize orthogonal communication resources to distinguish signals arriving from each client at the server, extract local models from the signals, and calculate their sum and average. However, this method has the disadvantage of low communication efficiency. An alternative method is to introduce the concept of over-the-air averaging \cite{yang2020federated,sery2021over} into the wireless FL framework. In simple terms, over-the-air averaging effectively utilizes the phenomenon where signals received from different clients are combined over the wireless medium, as shown in \eqref{eq:channel_model}. In a wireless MAC environment, the so-called over-the-air FL (OTA-FL) offers communication efficiency by eliminating the need for allocating orthogonal communication resources to each client, distinguishing it from conventional FL methods.

The first step of OTA-FL is to determine a receive power balancing parameter $\rho$. The parameter is determined by the server based on clients' CSI, and delivered to the clients before the first iteration of OTA-FL under the assumption that the channel distributions do not vary over time.\footnote{If the distribution of the channels changes as the iterations progress, the parameter should be updated before the start of each iteration.} Given the parameter $\rho$, in the $t$-th iteration of OTA-FL, the client $k$ designs a scaling factor $a_{k,t} > 0$ which satisfies
\begin{align}\label{eq:precoding}
\rho = a_{k,t} h_{k,t},\ \forall (k,t) \in \mathcal{K}\times\mathcal{T},
\end{align}
so that the powers of received gradients from the clients are equal at the server.
Next, the client transmits a signal $x_{k,t}=\sqrt{a_{k,t}}e^{-j\phi_{k,t}}g_{k,t}$ to the server, where $g_{k,t}$ is the local gradient model of the client $k$ in the $t$-th iteration. Then, according to \eqref{eq:channel_model}, the received signal at the server can be written as a scaled average gradient with additive noise:
\begin{align}\label{eq:average_gradient}
    y_t &= K \sqrt{\rho}  \underbrace{\frac{1}{K}\sum_{k\in\mathcal{K}}g_{k,t}}_{\text{average gradient}}+z_t.
\end{align}
Consequently, the average gradient can be used to update the global model.

In the context of OTA-FL, one inherent issue with this method lies in its susceptibility to the prevailing channel conditions between the server and the individual clients. Variations in signal strength received from clients can lead to a biased model aggregation, thereby compromising the accuracy of OTA-FL. Consequently, meticulous consideration of the parameter $\rho$ and the design of scaling factors $a_{k,t}$ for all $k\in\mathcal{K}$ and $t \in {1,2,\dots,\tau}$ is essential. However, ensuring the conditions stated in \eqref{eq:precoding} for all clients and time instances under a given $\rho$ typically demands higher transmission power to counter frequent occurrences of poor channels. Moreover, within scenarios of constrained transmission power, a situation extensively addressed in this article, selecting $\rho$ must align with the poorest client-to-server channel, inevitably impacting the overall OTA-FL performance.

To address these challenges, a channel threshold $h_{th} > 0$ was introduced to determine the quality of a channel \cite{seif2020wireless}\cite{chen2022decentralized}. In each OTA-FL iteration, only clients whose channel conditions with the server exceed the threshold will participate in model aggregation. The subset of clients with good channels participating in model aggregation in the $t$-th iteration is denoted as 
\begin{align}\label{eq:K_t}
\mathcal{K}_t = \{ k \mid k \in \mathcal{K} \text{ and } h_{k,t} \geq h_{th} \}
\end{align}
such that $\mathcal{K}_t \subset \mathcal{K}$ for all $t \in \{1,2,\dots,\tau\}$.
In this scenario, the average of local gradients in \eqref{eq:average_gradient} is computed over the set $\mathcal{K}_t$. Furthermore, since the channels of the clients participating in the model aggregation are now lower-bounded by $h_{th}$, and assuming that the entire transmission power $P$ is utilized to transmit the gradient when the channel is at $h_{th}$, $\rho$ can be chosen as described in \cite{seif2020wireless}\cite{chen2022decentralized}:
\begin{align}\label{eq:rho}
\rho = \frac{Ph_{th}}{W^2},
\end{align}
where $W$ represents the bounded norm of the gradient vectors.

\subsection{Differential Privacy and R\'enyi Differential Privacy}
To study the client privacy issues in OTA-FL, it is assumed that the server follows an honest-but-curious (HBC) adversary model\cite{liu2020privacy,seif2020wireless,Mohamed_amplification}. Here, an HBC server refers to a server that infers original data from local models received via uplink communication from clients while properly following the operational structure of OTA-FL. Since uplink communication between a client and the server is done $\tau$ times during the OTA-FL training, we aim to ensure privacy for all participating clients in all $\tau$ communications.

As mentioned earlier, differential privacy (DP) \cite{dwork2014algorithmic} is a standardized privacy framework used to protect sensitive data while preserving the utility of privatized data in various applications. Among various variants of DP, R\'enyi DP (RDP) \cite{mironov2017renyi}, in particular, offers efficiency in scenarios involving multiple iterations of information exchange, such as FL. Therefore, in this article, we  use RDP as the metric for privacy as in other FL studies in literature\cite{agarwal2021skellam}, \cite{wei2023personalized}.

Let $\mathcal{X}$ denote the space of sensitive data of interest. For adjacent databases, e.g., databases that have one datum difference, a mechanism must produce outputs that are nearly indistinguishable. Here, for fixed positive integers $\ell,m > 0$, such that $m>\ell$, we say that two databases $\mathbf{x},\mathbf{x}'\in \mathcal{X}^m$ are $\ell$-adjacent if they differ in $\ell$ entries. For a fixed positive integer $n >0$, let $\mathcal{Q}:\mathcal{X}^{m} \to \mathbb{R}^{n}$ be a query of which  sensitivity is defined as $\Delta = \lVert \mathcal{Q}(\mathbf{x})-\mathcal{Q}(\mathbf{x}')\rVert$. In the context of OTA-FL, the query $\mathcal{Q}$ refers to a function that outputs the average model update received from participating clients, which is computed over-the-air. The standard RDP of a randomized mechanism is formally defined as follows.

\begin{definition}
    \emph{(R\'enyi Differential Privacy)} Given a real number $\alpha \in (1,+\infty)$, a query $\mathcal{Q}:\mathcal{X}^{m} \to \mathbb{R}^{n}$, privacy parameter $\epsilon>0$, respectively, a randomized mechanism $\mathcal{M}:\mathbb{R}^n \to \mathbb{R}^n$ satisfies $(\alpha,\epsilon)$- RDP for all $\ell$-adjacent databases $\mathbf{x},\mathbf{x}'\in \mathcal{X}^m$: 
    \begin{align}
       D_{\alpha}[\mathcal{M}(\mathcal{Q}(\mathbf{x}))||\mathcal{M}(\mathcal{Q}(\mathbf{x}'))]\leq  \epsilon,
    \end{align}
    where $D_{\alpha}[P||Q]=\frac{1}{\alpha-1}\log \mathbf{E}[\left(P/Q\right)^{\alpha}]$ is the R\'enyi divergence.
\end{definition}
As mentioned, RDP is suitable as a privacy metric for FL because it possesses the following composition characteristics.
\begin{remark}\label{Remark1}
    \emph{(RDP Composition \cite{mironov2017renyi})} Suppose randomized mechanisms $\mathcal{M}_1$ and $\mathcal{M}_2$ achieve $(\alpha,\epsilon_1)$-RDP and $(\alpha,\epsilon_2)$-RDP, respectively. Then, the composite of the two mechanisms achieves $(\alpha,\epsilon_1+\epsilon_2)$-RDP.
\end{remark}
For instance in FL, if we designate the privacy protection mechanisms for the first and second global iterations as $\mathcal{M}_1$ and $\mathcal{M}_2$, and each achieves $(\alpha, \epsilon_1)$-RDP and $(\alpha, \epsilon_2)$-RDP respectively, it can be observed that the mechanism spanning both global iterations attains $(\alpha, \epsilon_1 + \epsilon_2)$-RDP.

In the meantime, throughout this article, we specifically focus on the Gaussian mechanism defined as follows.
\begin{definition}
    \emph{($\sigma^2$-Gaussian Mechanism)} For a query $\mathcal{Q}(\mathbf{x})$ of an input $\mathbf{x}$, the $\sigma^2$-Gaussian mechanism is defined as:
    \begin{align}
        \mathcal{M}_G(\mathcal{Q}(\mathbf{x}))\triangleq \mathcal{Q}(\mathbf{x})+ \mathcal{W},
    \end{align}
    where $\mathcal{W} \sim  \mathcal{N}(0,\sigma^2)$.
\end{definition}
The Gaussian mechanism simply operates by adding artificial Gaussian noise to the query. This aligns with scenarios in wireless communication, where it facilitates integrated noise analysis along with the additive white Gaussian noise experienced by the channel. Moreover, the Gaussian mechanism is known to achieve RDP, as mentioned in the following remark.
\begin{remark}\label{eq:Gaussian&RDP}
\emph{(Gaussian Mechanism and RDP \cite{mironov2017renyi})} For given $\alpha$ and  sensitivity $\Delta$, the $\sigma^2$-Gaussian mechanism achieves $\left(\alpha,\frac{\alpha \Delta^2}{2\sigma^2}\right)$-RDP.
\end{remark}
Therefore, taking these facts into account, in this paper, we will consider the Gaussian mechanism as a key enabler for enhancing the privacy of OTA-FL.

\section{Two Strategies for CwP in Over-the-Air Federated Learning}\label{sec:what clients with poor channels}
The previously discussed OTA-FL involves only selected clients participating in the learning process each iteration to achieve good learning performance. However, it is uncertain whether this approach will also have a positive effect on achieving high privacy performance. In this section, to address this question, we attempt a mathematical analysis of the learning convergence and RDP for two extreme strategies that CwP can adopt.

\subsection{Two Strategies for CwP}\label{private over-the-air model aggregation}
In the proposed OTA-FL framework, clients are categorized based on their channel conditions. Specifically, we classify clients with good channels (CwG) as those whose instantaneous channel gain exceeds a predetermined threshold, as defined in Section \ref{sec:OTA-FL}. These clients can reliably transmit gradient signals. Conversely, clients with poor channels (CwP) have channel gains below this threshold. Instead of transmitting gradient signals, they adopt alternative strategies, namely the \texttt{noisy} and \texttt{idle} methods.
Among these approaches, \texttt{idle} requires CwP clients to remain inactive during each iteration, similar to the basic OTA-FL framework. In contrast, \texttt{noisy} allows these clients to transmit artificial Gaussian noise. Since RDP generally provides stronger privacy guarantees with increased noise injection, replacing gradient signals with high-powered random noise can further enhance privacy. Building on this insight, we hypothesize that allowing CwP clients to transmit full powered random noise may improve the system’s overall performance. However, this does not imply that full-power noise transmission is always the optimal choice when considering both convergence error and privacy.
\begin{figure}
    \centering
    \includegraphics[width=0.45\textwidth]{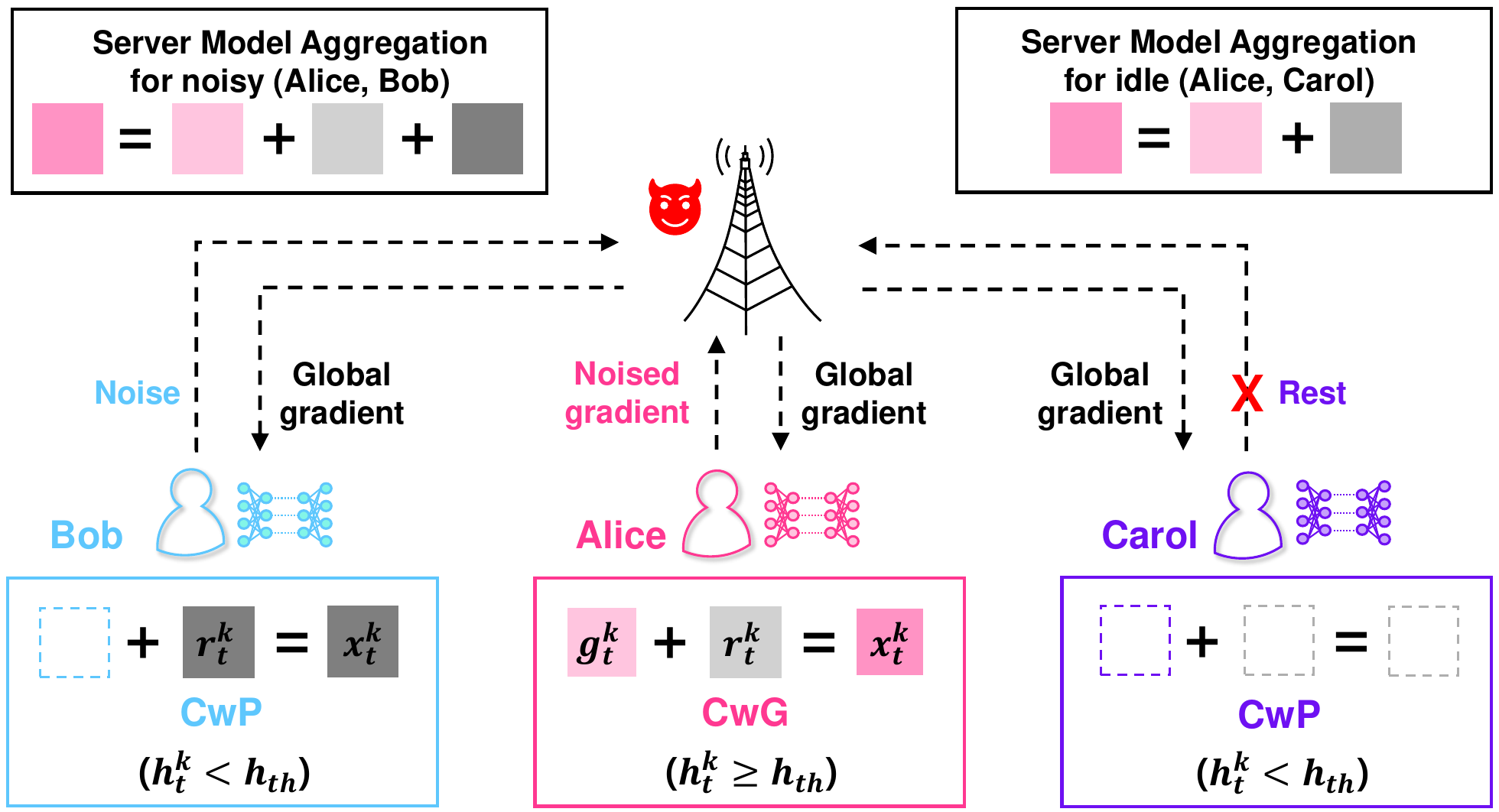}
    
    \caption{An illustration of \texttt{noisy} and \texttt{idle} strategies. Clients with good channels (CwG) transmit gradient and noise. However, clients with poor channels (CwP) choose either \texttt{noisy} or \texttt{idle} strategies, sending a noisy gradient or remaining idle.}
    \label{System_model}
    
\end{figure}
\subsubsection{\emph{\texttt{noisy}}}
The \texttt{noisy} CwP send artificial noise signals without gradient signals to enhance the OTA-FL privacy. The signal transmitted by the client $k$ in the $t$-th iteration can be written as: 
\begin{equation}
    x_{k,t}=
    \begin{cases}
    \sqrt{a_{k,t}}e^{-j\phi_{k,t}}(g_{k,t}+r_{k,t}),&  \mbox{if }  k\in\mathcal{K}_t\\
    r_{k,t},& \mbox{otherwise,}
    \end{cases}
\end{equation}
where $a_{k,t}\geq0$ is a scaling factor to satisfy the power condition  in \eqref{eq:precoding}, $g_{k,t}$ is a $d$-dimensional local gradient vector of client $k$, such that $g_{k,t} = \theta_{k,t}-\theta_{k,t-L}$, and $r_{k,t}\sim\mathcal{N}(0,\sigma_{r_{k,t}}^2I_d)$ is a $d$-dimensional artificial Gaussian noise vector so as to make its and other clients' signals private. To recall, as stated in \eqref{eq:K_t}, the set $\mathcal{K}_t$ represents the set of clients with channels surpassing the channel threshold $h_{th}$. All clients must properly design $a_{k,t}$ and $r_{k,t}$ to meet the power constraint:
\begin{equation}
    \begin{cases}\label{eq:power_constraint}
a_{k,t}\left[\|g_{k,t}\|^2+d\sigma_{r_{k,t}}^2\right] \leq P,&  \mbox{if }  k\in\mathcal{K}_t\\
    d\sigma_{r_{k,t}}^2=P,&   \mbox{otherwise,}
    \end{cases}
\end{equation}
where $P$ represents the available power of clients. Then, the received signal at the server can be expressed as
\begin{align}\label{update of case 1}
y_t =\sum_{k\in\mathcal{K}_t} \!\sqrt{\rho} g_{k,t} +\! \underbrace{\sum_{k\in\mathcal{K}_t}\sqrt{\rho}r_{k,t} +\!\sum_{k\not\in\mathcal{K}_t}\sqrt{h_{k,t}}r_{k,t}+z_t}_{\texttt{noisy}\text{ effective noise $q_t$}},
\end{align}
where $z_t \sim \mathcal{N}(0,\sigma_{z_t}^2)$ is an additive white Gaussian noise from the channel; $q_t\sim \mathcal{N}(0,\sigma_{q_t}^2)$ is an effective white Gaussian noise whose variance is 
\begin{align}\label{eq:noisy_var}
    \sigma_{q_t}^2=d\sum_{k\in\mathcal{K}_t} \rho \sigma_{r_{k,t}}^2+d\sum_{k\not\in\mathcal{K}_t} h_{k,t}\sigma_{r_{k,t}}^2+\sigma_{z_t}^2.
\end{align} 

\subsubsection{\emph{\texttt{idle}}}
The idle clients conserve energy and wait for better opportunities. The signal transmitted by the client $k$ in the $t$-th iteration can be expressed as
\begin{equation}
    x_{k,t}=
    \begin{cases}\sqrt{a_{k,t}}e^{-j\phi_{k,t}}(g_{k,t}+r_{k,t}),&  \mbox{if }  k\in\mathcal{K}_t \\
    0.&  \mbox{otherwise.}
    \end{cases}
\end{equation}
The clients in $\mathcal{K}_t$ must properly design $a_{k,t}$ and $r_{k,t}$ to satisfy the power constraint, given by:
\begin{equation}
    a_{k,t} \left(\|g_{k,t}\|^2+d\sigma_{r_{k,t}}^2\right) \leq P.
\end{equation}
The received signal at the server can be represented as
\begin{align}\label{update of case 2}
    y_t&=\sum_{k\in\mathcal{K}_t} \sqrt{\rho}g_{k,t} + \underbrace{\sum_{k\in\mathcal{K}_t} \sqrt{\rho}r_{k,t}+z_t}_{\texttt{idle}\text{ effective noise $q_t$}},
\end{align}
where the effective noise $q_t\sim \mathcal{N}(0,\sigma_{q_t}^2)$ is an additive white Gaussian noise whose variance is
\begin{align}\label{eq:idle_var}
\sigma_{q_t}^2=d \sum_{k\in\mathcal{K}_t} \rho\sigma_{r_{k,t}}^2+\sigma_{z_t}^2.
\end{align}

Upon receiving $y_t$, either \eqref{update of case 1} or \eqref{update of case 2}, the server performs global gradient estimation and obtains
\begin{align}\label{received_1}
    \hat{g}_t&=\frac{1}{\sqrt{\rho}K_t}y_t\\
    &=\frac{1}{K_t}\sum_{k\in\mathcal{K}_t}g_{k,t}
    +\frac{1}{\sqrt{\rho}K_t} q_t,
\end{align}
where $K_t = |\mathcal{K}_t|$ and 
\begin{align}\label{eq:sigma_qt}
     \sigma_{q_{t}}^2 = \begin{cases} \eqref{eq:noisy_var} & \text{for \texttt{noisy}}\\
     \eqref{eq:idle_var} & \text{for \texttt{idle}}
    \end{cases},
\end{align}
The OTA-FL with \texttt{noisy}/\texttt{idle} is described in detail in Algorithm \ref{algorithm2}. Note that in our framework, CwP are assumed to adopt one of these strategies exclusively throughout the training process.

\begin{algorithm}[t]
\small
\caption{OTA-FL with \texttt{noisy/idle}}\label{algorithm2}
\begin{algorithmic}
    \State{$\textbf{Initialize: }\theta_0$}
        \For {$t=0,1,2,...\tau$}
            \State{Broadcast global model $\theta_t$ to clients in $\mathcal{K}$.}
                \If {\texttt{noisy}}
                    \If{$k\in\mathcal{K}$}
                        \State{$\theta_{k,t,0}\gets\theta_t$}
                        \State{$\theta_{k,t,L}\gets$ \textbf{Local update} ($\theta_{k,t,0},\mathcal{D}_k$)}
                        \State{Get $g_{k,t}=\theta_{k,t,L}-\theta_{k,t,0}$}
                        \State{Transmit $\sqrt{a_{k,t}}e^{-j\phi_{k,t}}(g_{k,t}+r_{k,t})$ to the server.}
                    \Else
                        \State{Transmit $r_{k,t}$ to the server.}
                    \EndIf
            \State$\theta_{t+1}=\theta_t-\eta_t \left(\frac{1}{K_t}\sum_{k\in\mathcal{K}_t}\sum_{\ell=1}^L\nabla{f_k(\theta_{k,t,\ell})}+\frac{q_t}{K_t}\right),$ where $\theta_t=\frac{1}{K_t} \sum_{k\in \mathcal{K}_t}\theta_{k,t}$ and $q_t$ is defined in \eqref{update of case 1}.
                \ElsIf {\texttt{idle}}
                    \If{$k\in\mathcal{K}$}
                        \State{$\theta_{k,t,0}\gets\theta_t$}
                        \State{$\theta_{k,t,L}\gets$ \textbf{Local update} ($\theta_{k,t,0},\mathcal{D}_k$)}
                        \State{Get $g_{k,t}=\theta_{k,t,L}-\theta_{k,t,0}$}
                        \State{Transmit $\sqrt{a_{k,t}}e^{-j\phi_{k,t}}(g_{k,t}+r_{k,t})$ to the server.}
                    \EndIf
                \EndIf
            \State$\theta_{t+1}= \theta_t -\eta_t \left(\frac{1}{K_t}\sum_{k\in\mathcal{K}_t}\sum_{\ell=1}^L\nabla{f_k(\theta_{k,t,\ell})}+\frac{q_t}{K_t}\right)$
            where $\theta_t=\frac{1}{K_t} \sum_{k\in \mathcal{K}_t}\theta_{k,t}$ and $q_t$ is defined in \eqref{update of case 2}.
        \EndFor
    \\
    \State{\textbf{Local update} $(\theta_{k,t},\mathcal{D}_k)\textbf{:}$}
        \For{$l=1,...,L$}
        \State{$\theta_{k,t,\ell}=\theta_{k,t,\ell-1}-\eta_t\nabla{f_k(\theta_{k,t,\ell-1})}$}
        \EndFor
        \State{Return $\theta_{k,t,L}$}
\end{algorithmic}
\end{algorithm}

\subsection{R\'enyi Differential Privacy of OTA-FL with \emph{\texttt{noisy/idle}}} \label{newDP}
To explore RDP in OTA-FL with \texttt{noisy}/\texttt{idle}, it is assumed that the local gradients are confined within a positive real number denoted as $W > 0$. In addition, the channel connecting the server and client is presumed to be independently and identically distributed (i.i.d.) over time and across clients. For the sake of simplicity, let $p$ be the probability of an i.i.d. channel exceeding the threshold $h_{th}$. Subsequently, the following theorem is established.

\begin{theorem}\label{renyi_theorem}
     For given $\alpha\geq2$, $0\leq p\leq1$, $W>0$ and $\sigma_{q_{t}}^2$ as \eqref{eq:sigma_qt}, the total $\tau$ global iterations of OTA-FL achieves $\left(\alpha, \epsilon\right)$-RDP, where $\epsilon = \frac{\tau\log 2}{\alpha-1}+ \frac{\tau\alpha}{\alpha-1}\log\left(pe^{\frac{(\alpha-1)W^2}{\sigma_{q_t}^2}}+1\right)$.
\end{theorem}
\begin{IEEEproof}
A detailed is provided in Appendix \ref{AppendixA}.
\end{IEEEproof}

Theorem \ref{renyi_theorem} reveals that the privacy guarantee of the OTA-FL with \texttt{noisy}/\texttt{idle}, represented by $\epsilon$, is critically influenced by two key parameters: the probability of participation $p$ and the noise variance $\sigma^2_{q_t}$. 
Specifically, an increment in the probability of participation leads to a greater number of clients participating in gradient signal transmission, which, in turn, increases the potential risk of privacy breaches. On the other hand, an increase in the noise variance introduces more noise into the gradient signals, serving as an effective countermeasure against privacy invasion.

\subsection{Convergence Analysis for OTA-FL with \emph{\texttt{noisy}}/\emph{\texttt{idle}}}\label{convergence}
For theoretical analysis of the OTA-FL with \texttt{noisy}/\texttt{idle}, we make the following assumptions which are typically made in analyzing FL family \cite{li2019convergence,9252927,bottou2018optimization,jhunjhunwala2022fedvarp}.

\noindent\emph{{A1)} The global loss function $f$: $\mathbb{R^d}\rightarrow\mathbb{R}$ is $M$-smooth, namely, for all $\theta_1,\theta_2\in\mathbb{R}^d$, it holds that} 
\begin{align}
    f(\theta_1)-f(\theta_2)\leq(\theta_1-\theta_2)^T\nabla f(\theta_2)+\frac{1}{2}M\lVert\theta_1-\theta_2\rVert^2.
\end{align}

\noindent\emph{{A2)} The global loss function $f$ is $\mu$-strongly convex with constant $\mu>0$, that is for all $\theta_1,\theta_2 \in\mathbb{R}^d$:}
\begin{align}
    f(\theta_1)-f(\theta_2)\geq(\theta_1-\theta_2)^{T}\nabla f(\theta_2)+\frac{1}{2}\mu\lVert\theta_1-\theta_2\rVert^2,
\end{align}

\noindent\emph{A3)} For every client $k\!\in\!\{1,\dots,K\}$, any model parameter $\theta\!\in\!\mathbb{R}^d$,   
and a mini-batch $\xi_k$ drawn from the local dataset $\mathcal D_k$, $\mathbf{E}_{\xi_k}\!\bigl[\nabla f_k(\theta;\,\xi_k)\bigr] = \nabla f_k(\theta)$ and it satisfies $\mathbf{E}[\lVert\nabla f_k(\theta;\,\xi_k)\rVert^2]\leq G^2$, with a fixed constant $G>0$.

\noindent{\emph{A4)}} There exists a constant $\sigma_G\!\ge 0$ such that for all $\theta\in\mathbb{R}^d$, $\frac{1}{K}\sum_{k=1}^K\|\nabla f_k(\theta)-\nabla f(\theta)\|^2
    \le \sigma_G^{2}.$

\begin{theorem}\label{convergence_theorem}
Consider a scenario where the global gradient is derived via OTA-FL with \emph{\texttt{noisy}}/\emph{\texttt{idle}} with a step size of $\eta_t=\frac{4}{\mu(a+t)}$, where $a > \max\{8bM,(\sqrt{2}+1)M\}$. Assuming that conditions {A1}, {A2}, {A3} and {A4} are met and that the locally updated gradients remain bounded, the convergence error is upper-bounded as 
\begin{align}
\mathbf{E}[f(\Theta_{\tau})]-f(\theta^*) \leq \gamma,
\end{align}
where $\gamma = \frac{\mu a^3}{4S_\tau}\mathbf{E}[\lVert\theta_0-\theta^*\rVert^2] + \frac{2\tau(\tau + a)}{\mu S_\tau} (\frac{4L^2G^2}{K_t}  +  \frac{\sigma_{q_t}^2}{K_t^2\rho}+L^4M^2\eta_t^2G^2+3\sigma_G^2)$, $\Theta_{\tau}=\frac{1}{S_\tau}\sum_{t=0}^{\tau-1}\beta_t\theta_{t}$, $S_\tau=\sum_{t=0}^{\tau-1}\beta_{t}\geq\frac{1}{3}{\tau}^3$ and $\beta_t=(a+t)^2$.
\end{theorem}
\begin{IEEEproof}
A detailed proof is provided in Appendix \ref{AppendixB}.
\end{IEEEproof}
Theorem \ref{convergence_theorem} is derived under the standard assumptions of both strong convexity (A2) and uniformly bounded gradients (A3), both of which are widely used in the FL literature to obtain tractable convergence results \cite{mao2024leveraging,yan2023over,xue2023over,li2019convergence,sery2021over,stich2018local}.

In the meantime, as shown in \cite{nguyen2019new}, when both  (A2) and (A3) hold, the following inequality must be satisfied: 
\begin{align}
    f(\theta_0)-f(\theta^*)\leq \frac{G^2}{2\mu}.
\end{align}
By definition of strong convexity, it further implies
\begin{align}\label{eq: contradiction}
    (\theta_0-\theta^*)^T\nabla f(\theta^*)+\frac{\mu}{2}\|\theta_0-\theta^*\|^2\leq\frac{G^2}{2\mu}.
\end{align}
If inequality \eqref{eq: contradiction} does not hold, then (A2) and (A3) cannot be applied at the same time. \eqref{eq: contradiction} is satisfied either when the bound on the gradient norm $G$ is sufficiently large relative to the distance between the initial model and the obtained model, or when the initial model is already close to the obtained model. Such cases often occur in fine-tuning scenarios or when the privacy requirements are very stringent, leading to only minor adjustments during training. Recognizing that the aforementioned conditions may not hold in more general scenarios, we have extended our convergence analysis to cover the non-convex case as well.
\begin{theorem}\label{convergence_theorem_nonconvex}
    Consider the same scenario as in Theorem \ref{convergence_theorem}, but without assuming strong convexity (i.e., assuming only A1, A3 and A4 hold). Then, the cumulative expected squared norm is upper-bounded as
    \begin{align}
    \frac{1}{\tau}\sum_{t=0}^{\tau-1}&\mathbf{E}\left[\|\nabla f(\theta_t)\|^2\right] \leq \gamma, 
\end{align}
where $\gamma=\sum_{t=0}^{\tau-1}\eta_t^2\frac{L^3}{3\tau}M^2G^2  + \frac{2}{\tau}[f(\theta_0)-\theta^*]
    \!+\!\frac{M}{\tau}\sum_{t=0}^{\tau-1}\eta_t\left(\frac{4L^2G^2}{K_t} \!+\! \frac{\sigma_{q_t}^2}{K_t^2\rho} \!+\! L^4M^2\eta_t^2G^2+3\sigma_G^2+3G^2\right).$
\end{theorem}
\begin{IEEEproof}
    A detailed proof is provided in Appendix \ref{AppendixC}.
\end{IEEEproof}

The term $\gamma$ quantifies the expected value of the loss function evaluated at the averaged model parameters after $\tau$ global iterations. It reflects the expected performance of the aggregated global model, incorporating contributions from multiple local models across the network. The convergence of OTA-FL with \texttt{noisy}/\texttt{idle} is also mainly affected by the parameter $\rho$. To be specific, in $\frac{4L^2G^2}{K_t}+\frac{\sigma_{q_t}^2}{K_t^2\rho}$, the total number of clients $K_t$ and noise variance $\sigma_{q_t}^2$ are dependent on $\rho$. An increase in $K_t$ means more clients transmitting gradient signals, lowering the convergence error. Conversely, a higher noise variance $\sigma_{q_t}^2$ results in more perturbed gradient signals, thereby escalating the convergence error.

\section{Client-Driven Power Balancing for Over-the-Air Federated Learning}\label{sec:client driven power balancing}

\begin{figure}[!t]
    \centering
    \subfigure[]{\includegraphics[width=0.48\linewidth]{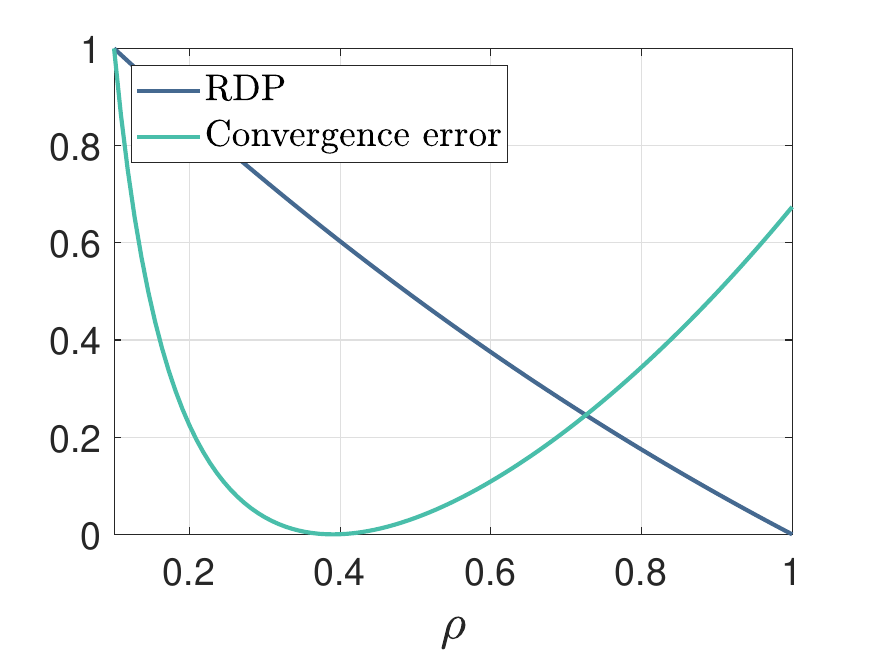}}
    \vspace{-2mm}
    \subfigure[]{\includegraphics[width=0.48\linewidth]{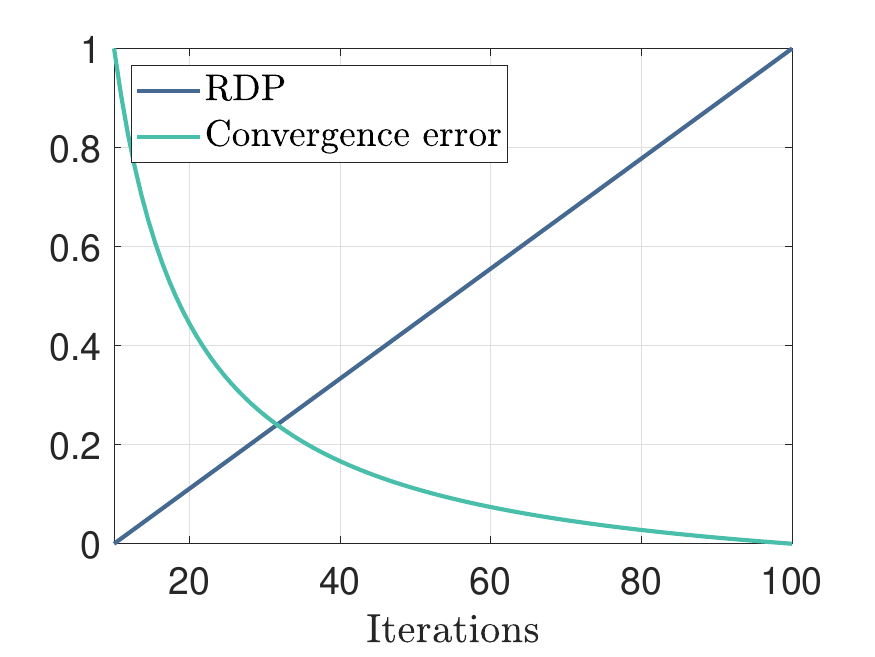}}
    
    \caption{Comparison of convergence error and RDP with respect to (a) $\rho$ and (b) the global iterations.}
    \label{fig:trade-offs}
    
\end{figure}

In the previous section, we formulated analytical results using two strategies for CwP. In this section, we propose the \textit{client-driven power balancing} (\texttt{CDPB}) strategy. As previously outlined, the convergence error, denoted as $\gamma$, and the RDP, denoted as $\epsilon$, are influenced by the number of CwG $K_t$ as well as the noise $\sigma_{q_t}^2$. These parameters can be determined through the power balancing parameter $\rho$. Given that $K_t$ and $\sigma_{q_t}^2$ are closely related to the channel distribution of each client, the server can utilize this channel distribution to determine the appropriate balancing parameter $\rho$, thereby balancing learning efficiency and privacy.

Once $\rho$ is determined, clients can perform power balancing without requiring CSI feedback to the server. This is a more realistic scenario, as clients are inclined to conceal their gradient information from the server and independently manage the transmission of gradient signals and power balancing. Additionally, as the number of global iterations $\tau$ increases, the convergence error $\gamma$ tends to decrease, but concerns regarding $\epsilon$ escalate due to the increased exposure of client gradient signals to the server. These observations complicate the training process in OTA-FL. Fig.  \ref{fig:trade-offs} illustrates these observations. 

Thus, our primary objective is to minimize the convergence error while maintaining a robust level of privacy. To effectively balance convergence error and privacy, we introduce a utility function $\mathcal{G}(\rho,\tau)$, which quantifies the privacy gain against a specific convergence error after $\tau$ iterations:
\begin{align}
\mathcal{G}(\rho,\tau) = \lambda_1\gamma + \lambda_2 \epsilon,
\end{align}
where $\lambda_1$ and $\lambda_2$ are constant parameters that modulate the respective contributions to the convergence error and achievable privacy.

Consider an optimization problem\footnote{If one wishes to optimize the artificial noise level more finely, an extra design variable such as a scaling factor $\xi\in[0,1]$ could be introduced so that the effective noise power becomes $\xi\sigma_{n}^2$, where $\sigma_{n}^2$ is the maximum noise power transmitted by CwP. This would enable continuous tuning of the noise level and provide a richer trade-off between privacy and convergence. However, in this paper,  we focus on two extreme strategies, \texttt{noisy} and \texttt{idle} to clearly delineate and compare the trade-offs between privacy protection and convergence performance without the additional complexity of optimizing $\xi$.}
\begin{subequations}\label{OP:P1}
\begin{align}
\mathcal{P}_1: \underset{\rho,\tau}{\text{min}} ~~ &\mathcal{G}(\rho,\tau)\label{ob:p1} \\
\text{s.t.} ~~ &\gamma \leq \bar{\gamma},\label{ct:gamma}\\
&\epsilon \leq \bar{\epsilon},\label{ct:epsilon}\\
&\sigma_{q_t}^2 \leq 2KP+\sigma_{z_t}^2,\label{ct:noise}\\
&0 \leq \rho \leq \frac{P}{W^2},\label{ct:rho}
\end{align}
\end{subequations}
where $\bar{\gamma}~ ( >0)$ and $\bar{\epsilon}~ (> 0)$ in the first two thresholds are the acceptable convergence error and privacy requirement, respectively and the last two constraints represent a feasible range of received noise variance $\sigma_{q_t}^2$ and receive power balancing parameter $\rho$, respectively. Constraints \eqref{ct:gamma} and \eqref{ct:epsilon} represent the system’s minimum quality-of-service (QoS) requirements. If these thresholds are set too stringently, making it impossible to satisfy both constraints simultaneously, the minimum QoS cannot be achieved. In such cases, we declare the feasible set empty, preventing the initiation of the training process that follows the optimization stage. Conversely, if the objective is to find an optimal solution regardless of the QoS requirements, we can gradually increase these threshold values until the feasible set becomes non-empty, at which point we apply the proposed algorithm.

The objective function in $\mathcal{P}_1$ involves the global iteration parameter $\tau$, resulting in a mixed-integer programming (MIP) problem, which is inherently complex. To address this efficiently, we decompose the problem into two subproblems and solve them sequentially. In the first stage, we propose a straightforward approach by precisely narrowing down the search space to a feasible set $\mathcal{T} \subseteq \mathbb{Z}^{+}$ that satisfies thresholds \eqref{ct:gamma} and \eqref{ct:epsilon} for all feasible $\rho$. In the second stage, for each given $\tau\in\mathcal{T}$, we can omit thresholds \eqref{ct:gamma} and \eqref{ct:epsilon}, allowing us to fully concentrate on finding the values of $\rho$ and $\tau$ that minimize $\mathcal{G}(\rho, \tau)$.

\subsection{Stage I - Obtaining the Feasible Set $\mathcal{T}$}
Stage I aims to determine the feasible set $\mathcal{T}$ that satisfies the thresholds \eqref{ct:gamma} and \eqref{ct:epsilon}. Since $\gamma$ and $\epsilon$ are influenced by both $\rho$ and $\tau$, we need to determine $\tau$ for all possible values of $\rho$. Given that $\bar{\gamma}$ and $\bar{\epsilon}$ only affect $\gamma$ and $\epsilon$, respectively, we find the sets $\mathcal{T}_{\gamma}$ using $\bar{\gamma}$ and $\mathcal{T}_{\epsilon}$ using $\bar{\epsilon}$ separately and then derive $\mathcal{T}$ by taking the intersection of $\mathcal{T}_{\gamma}$ and $\mathcal{T}_{\epsilon}$.

Before calculating $\mathcal{T}_{\gamma}$ and $\mathcal{T}_{\epsilon}$, we examine the characteristics of the $\gamma$ and $\epsilon$ functions. Since $\gamma$ is inversely proportional to $\tau$, the maximum value of $\tau$ in $\mathcal{T}_{\gamma}$ is $\tau_{\gamma,\text{max}} = \infty$, at which point $\gamma = 0$. This implies that as the number of global iterations approaches infinity, the convergence error converges to zero. Similarly, since the $\epsilon$ function is linearly proportional to $\tau$, the minimum value of $\tau$ in $\mathcal{T}_{\epsilon}$ is $\tau_{\epsilon,\text{min}} = 0$, at which point $\epsilon = 0$. This implies that when clients do not communicate with the server, privacy is maximally guaranteed.

To obtain $\mathcal{T}_{\gamma}$, we only need to determine $\tau_{\gamma,\text{min}}$, resulting in $\mathcal{T}_{\gamma} = \{\tau_{\gamma,\text{min}}, \tau_{\gamma,\text{min}}+1, \ldots\}$. Similarly, to obtain $\mathcal{T}_{\epsilon}$, we only need to determine $\tau_{\epsilon,\text{max}}$, resulting in $\mathcal{T}_{\epsilon} = \{0, 1, \ldots, \tau_{\epsilon,\text{max}}\}$. Finally, $\mathcal{T}$ can be represented as $\mathcal{T} = \{\tau_{\gamma,\text{min}}, \tau_{\gamma,\text{min}}+1, \ldots, \tau_{\epsilon,\text{max}}\}$.

Now, we focus on determining $\tau_{\gamma,\text{min}}$, which can be obtained by solving the following subproblem:
\begin{subequations}\label{OP:P2}
\begin{align}
\mathcal{P}_2:  &\min_{\rho}  \tau  \\
\text{s.t.} & \ \  \gamma \leq \bar{\gamma},\\
& \ \  \sigma_{q_t}^2 \leq 2KP+\sigma_{z_t}^2,\\
& \ \  0 \leq \rho \leq \frac{P}{W^2}.
\end{align}
\end{subequations}
\begin{lemma}\label{lemma:1}
    In Stage I, under Rayleigh fading and given the acceptable convergence error $\bar{\gamma}$, $\tau_{\gamma,\text{min}}$ can be obtained as follows:
        \begin{align}
        \tau_{\gamma,\text{min}}=\left\lceil\frac{1}{\bar{\gamma}}\left(\frac{4L^2G^2}{K_t}+\frac{\sigma_{q_t}^2}{K_t^2\rho_{\gamma}}\right)\right\rceil,
    \end{align}
    where $\rho_{\gamma}$ is 
        \begin{itemize}
        \item For \texttt{CDPB-n},
            \begin{align}\label{eq:T_min_noisy}
                \rho_{\gamma} = \frac{P\sigma^2(\sqrt{4a_n+9}-1)}{W^2(a_n+2)},
            \end{align}
            where $a_n=4M^2G^2-W^2$.
        \item For \texttt{CDPB-i}, 
            \begin{align}\label{eq:T_min_idle}
                \rho_{\gamma} = \frac{P\sigma^2(\sqrt{4a_i+1}-1)}{a_iW^2},
            \end{align}
            where $a_i=\frac{4L^2G^2}{W^2}$.
    \end{itemize}
\end{lemma}
\begin{IEEEproof}
    A detailed proof is provided in Appendix \ref{AppendixD}.
\end{IEEEproof}

From Lemma \ref{lemma:1}, we can determine $\tau_{\gamma,\text{min}}$ using a fixed $\rho_\gamma$. Similarly, $\tau_{\epsilon,\text{min}}$ can be obtained by solving the following subproblem:
\begin{subequations}\label{OP:P3}
\begin{align}
\mathcal{P}_3:  &\max_{\rho} \tau\\
\text{s.t.} & \ \  \epsilon \leq \bar{\epsilon}, \\
& \ \  \sigma_{q_t}^2 \leq 2KP+\sigma_{z_t}^2,\\
& \ \  0 \leq \rho \leq \frac{P}{W^2}.
\end{align}
\end{subequations}

\begin{lemma}\label{lemma:2}
    In Stage I, given the privacy requirement $\bar{\epsilon}$, the $\tau_{\epsilon,\text{max}}$ can be obtained as:
    \begin{align}\label{eq:T_max}
        \tau_{\epsilon,\text{max}}=\left\lfloor\frac{\bar{\epsilon}\log 2}{\alpha-1}+ \frac{\bar{\epsilon}\alpha}{\alpha-1}\log\left(p e^{\frac{(\alpha-1)W^2}{\sigma_{q_t}^2}}+1\right)\right\rfloor
    \end{align}
    where $\rho_{\epsilon}=\frac{P}{W^2}$ for both \texttt{CDPB-n} and \texttt{CDPB-i}.
\end{lemma}
\begin{IEEEproof}
    A detailed proof is provided in Appendix \ref{AppendixE}. 
\end{IEEEproof}
From Lemma 1 and 2, we can finally derive a feasible set $\mathcal{T}=\{\tau_{\gamma, \text{min}}, \tau_{\gamma, \text{min}}+1,...,\tau_{\epsilon, \text{max}}\}$.

\subsection{Stage II - Obtaining  $\rho_{opt}$ and $\tau_{opt}$}
Based on Stage I, we can derive a feasible set $\mathcal{T}$ that satisfies acceptable convergence error and privacy requirements for all possible $\rho$ values. We then calculate the utility function $\mathcal{G}(\rho, \tau)$ for a finite number of values, specifically $|\mathcal{T}|$ values, each corresponding to a feasible value of $\tau \in \mathcal{T}$. Given that the utility function $\mathcal{G}(\rho, \tau)$ is convex, there exists a unique value of $\rho$ that minimizes $\mathcal{G}(\rho, \tau)$ for each feasible $\tau$. Therefore, to find the minimum value of the utility function $\mathcal{G}(\rho, \tau)$ over all feasible pairs $(\rho, \tau)$, we perform a line search over the $|\mathcal{T}|$ feasible points. In other words, the optimization problem can be solved by evaluating the utility function at each of the $|\mathcal{T}|$ feasible $(\rho, \tau)$ pairs and selecting the pair that yields the minimum utility value. Consequently, $\mathcal{P}_1$ transforms into solving $\mathcal{P}_4$ as follows: 
\begin{subequations}\label{OP:P4}
\begin{align}
\mathcal{P}_4: &\min_{\rho,\tau} \mathcal{G}(\rho,\tau) \\
\text{s.t.} & \ \ \tau\in\mathcal{T},\\
& \ \ \sigma_{q_t}^2 \leq 2KP+\sigma_{z_t}^2,\\
& \ \ 0 \leq \rho \leq \frac{P}{W^2}.
\end{align}
\end{subequations}

Since the terms in $\mathcal{G}(\rho, \tau)$ are convex with respect to $\rho$, we can easily find the optimal $\rho$ for every element in $\mathcal{T}$. For \texttt{CDPB-i}, we can also find the exact solution for $\rho$, which is:

\begin{theorem}\label{theorem:rho}
    Given the global iteration $\tau$, under the conditions $\rho < \frac{P}{W^2}$, $\alpha \geq 2$, and Rayleigh channel distribution, $\rho$ can be obtained as:
    \begin{align}\label{eq:rho_opt}
        \rho_{\tau} = \frac{2P\sigma^2}{W^2}\sqrt{\frac{a}{\frac{W^2(\alpha-1)}{2P\sigma^2}+b-1}},
    \end{align}
    where $a = \frac{\lambda_1}{\lambda_2 \tau^2K_tW^2}$ and $b = \frac{4\lambda_1 L^2G^2}{\lambda_2 \tau^2K_t}$.
\end{theorem}

\begin{IEEEproof}
    The proof is similar to that of Lemma \ref{lemma:1}.
\end{IEEEproof}

On the other hand, for \texttt{CDPB-n}, the problem can be solved using numerical methods such as the bisection method, as shown in Algorithm \ref{Bisection Method for noisy}, based on the Lagrange multiplier method. Finally, by comparing $\mathcal{G}(\rho_{\tau},\tau)$, which can be derived from Theorem \ref{theorem:rho} or Algorithm \ref{Bisection Method for noisy}, for every $\tau \in \mathcal{T}$, we can determine the optimal $\rho_{\text{opt}}$ and $\tau_{\text{opt}}$ corresponding to the minimum utility function $\mathcal{G}(\rho, \tau)$. Once $\rho_{\text{opt}}$ and $\tau_{\text{opt}}$ are computed, clients and the server can begin training with the power balancing parameter $\rho_{\text{opt}}$ for $\tau_{\text{opt}}$ global iterations. During training, clients do not need to report their CSI to the server but can independently perform power balancing.

\subsection{Complexity Analysis}
We now analyze the computational complexity of our two-stage optimization algorithm. In Stage I, the feasible set $\mathcal{T}$ is computed using basic arithmetic through equations in Lemma 1, so its cost is negligible. In Stage II, we perform a line search over $\mathcal{T}$ to determine the optimal parameters. In particular for \texttt{CDPB-i}, where $\rho$ is computed via a closed-form expression, the per-candidate cost is $\mathcal{O}(1)$, leading to an overall complexity of $\mathcal{O}(\mathcal{|T|})$. On the other hand, for \texttt{CDPB-n}, $\rho$ is determined using a bisection method with a per-candidate cost of $\mathcal{O}(\log(1/\Psi))$ where $\Psi$ is tolerance, resulting in an overall complexity of $\mathcal{O}(|\mathcal{T}|\log(1/\Psi))$.

\begin{algorithm}[t]
\small
\caption{Bisection Method for \texttt{CDPB-n}}
\label{Bisection Method for noisy}
\begin{algorithmic}[1]
\State \textbf{Input: }Lagrange multiplier $f(x)$, interval $[0, P]$, tolerance $\Psi$, maximum iterations $N_{\text{max}}$
\State Initialize $n \gets 0, a \gets 0, b \gets P, c\gets0$ 
\While{$n < N_{\text{max}}$}
    \State Update $c \gets (a + b)/2$
    \If{$f(c) = 0$ or $(b - a)/2 < \Psi$}
    \Else
    \State Update $n \leftarrow n + 1$
    \If{$\text{sign}(f(c)) = \text{sign}(f(a))$}
        \State Update $a \leftarrow c$
    \Else
        \State Update $b \leftarrow c$
    \EndIf
    \EndIf
\EndWhile\\
\Return $c$
\end{algorithmic}
\end{algorithm}

\begin{algorithm}[t]
\small
\caption{Two-Stage Optimization of \texttt{CDPB}}
\label{two-stage_power_control}
\begin{algorithmic}[1]

\State \textbf{Stage 1: Obtain $\mathcal{T}$}
\State Initialize $\mathcal{T}\gets\emptyset$ 
\If{\texttt{CDPB-n}}
    \State{Obtain $\mathcal{T}$ using Eq. \eqref{eq:T_min_noisy} and \eqref{eq:T_max}}
\Else  
    \State{Obtain $\mathcal{T}$ using Eq. \eqref{eq:T_min_idle} and \eqref{eq:T_max}}
\EndIf
\State \textbf{return} $\mathcal{T}$ 
\State \textbf{Stage 2: Obtain $\tau_{opt}, \rho_{opt}$}
\State {Initialize $\rho_{opt}$ $\gets$ 0, $\tau_{opt}$ $\gets$ 0} 
\For{each $\tau$ in $\mathcal{T}$}
    \If{\texttt{CDPB-n}}
    \State{Obtain $\rho_{\tau}$ using Algorithm \ref{Bisection Method for noisy}}
    \Else  
    \State{Obtain $\rho_{\tau}$ using Eq. \eqref{eq:rho_opt}}
    \EndIf
    \If{$\tau=\lceil \tau_{\text{min}} \rceil$}
        \State Update $\rho_{opt}\gets\rho_{\tau}$
        \State Update $\tau_{opt}\gets\lceil \tau_{\text{min}} \rceil$
        \State Update $\mathcal{G}(\rho_{opt},\tau_{opt})\gets\mathcal{G}(\rho_{opt},\lceil \tau_{\text{min}} \rceil)$
    \Else
        \If{$\mathcal{G}(\rho, \tau)<G_{opt}$}
        \State Update $\rho_{opt}\gets\rho_{\tau}$ 
        \State Update $\tau_{opt}\gets\tau_{\tau}$ 
        \EndIf
    \EndIf
\EndFor
\State \textbf{return} $\rho_{opt}$ and $\tau_{opt}$ \Comment{Return optimal values}
\end{algorithmic}
\end{algorithm}

\section{Experimental Results}\label{sec:experimental results}
In this section, we show the experimental results that corroborate the efficacy of the proposed \texttt{CDPB}. Software codes has been implemented in Python 3.8 and executed on an Ubuntu server equipped with NVIDIA GeForce RTX 4090 GPUs. 
The default experiment settings are given as follows. We evaluate our approach using an optimized CNN for CIFAR-100 classification. The network comprises approximately 0.84 million parameters and features a lightweight design with a feature extractor consisting of three convolutional blocks followed by a two-layer fully connected classifier. The CIFAR-100 dataset contains 60,000 32×32 color images distributed among 100 classes (600 images per class), with 50,000 images allocated for training and 10,000 for testing. These classes are further grouped into 20 superclasses, each comprising five related classes that span a wide range of objects.

Training is performed using SGD with an initial learning rate of 0.05 and a cosine annealing scheduler. In our simulation, 100 clients participate, each conducting 5 local iterations before aggregation. $\lambda_1$ and $\lambda_2$ are set as $1$ and $1\times10^{-5}$, respectively. The $\bar{\gamma}$ and $\bar{\epsilon}$ are set $1\times10^{-2}$ and $100$, respectively. We compare the proposed scheme against the following baseline power balancing approaches.
\begin{itemize}
    \item \texttt{CDPB}: This is our proposed power balancing strategy. It focuses on minimizing $\mathcal{G}(\rho, \tau)$ considering both the acceptable convergence error $\bar{\gamma}$ and the privacy requirement $\bar{\epsilon}$. Note that it is designed to allow clients to autonomously perform power balancing based on $\rho_{opt}^*$, which is obtained through channel distribution.
    \item \texttt{$\bar{\gamma}$ based}: This power balancing strategy focuses on meeting the acceptable convergence error $\bar{\gamma}$. Note that it does not consider privacy requirement.
    \item \texttt{$h_{\text{min}}$ based}\cite{liu2020privacy}: This strategy requires all clients to transmit their gradient signals to the server at every global iteration. Note that the performance is determined by the channel gain of the client with the worst channel conditions. Specifically, the $\rho$ is set as $\rho=\frac{Ph_{\text{min}}}{W^2}$. 
    \item \texttt{Independent sampling (IS) based}\cite{Mohamed_amplification}: This strategy requires all clients to perform sampling without accounting for channel distribution information. However, the server can access each client's CSI to determine power balancing and identify which clients are participating in each iteration.
    \item \texttt{Noise-Free}: This strategy is the same as the proposed power balancing strategy but assumes a noise-free scenario. Specifically, problems are solved with $\sigma_{r_k,t}^2=0$.
\end{itemize}

\begin{figure}
    \centering
    \includegraphics[width=0.48\textwidth]{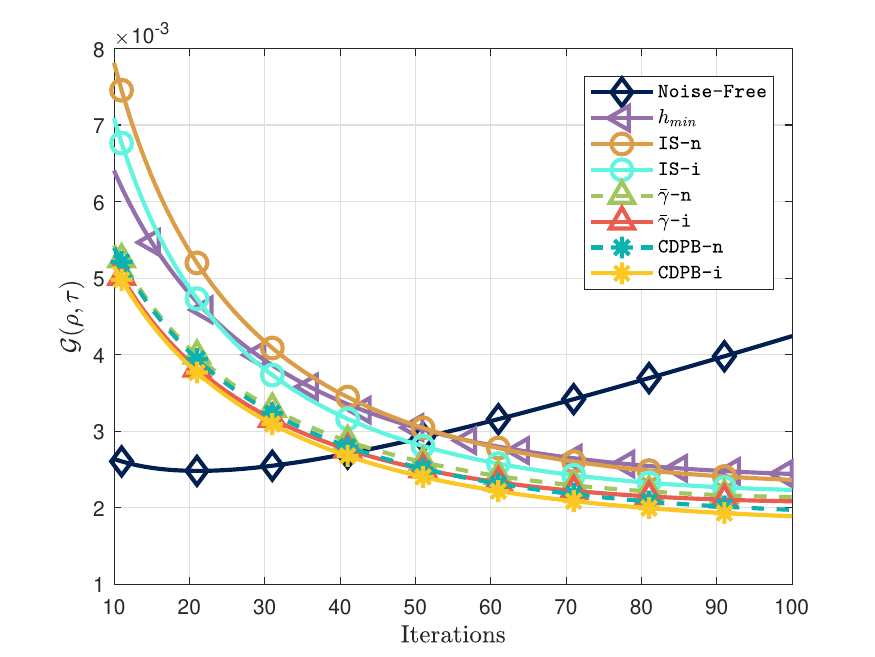}
    \caption{$\mathcal{G}(\rho, \tau)$ under different power balancing strategies.}
    \label{fig:G_T}
\end{figure}

\textbf{Achievable $\mathcal{G}(\rho,\tau)$} \quad
Fig. \ref{fig:G_T} illustrates the guaranteed values of $\mathcal{G}(\rho, \tau)$ under various power balancing strategies. As iterations progress, all approaches except \texttt{Noise-Free} show a decreasing trend, with \texttt{CDPB} consistently outperforming the others across all ranges. However, in the early iterations, \texttt{Noise-Free} achieves better performance than any other strategy. This is because \texttt{Noise-Free} avoids adding artificial noise to maintain RDP, leading to substantial improvements in convergence error at the beginning. Yet, as iterations increase, the $\mathcal{G}(\rho, \tau)$ for \texttt{Noise-Free} rises sharply due to the loss of RDP.

\begin{figure}
    \centering    \includegraphics[width=0.49\textwidth]{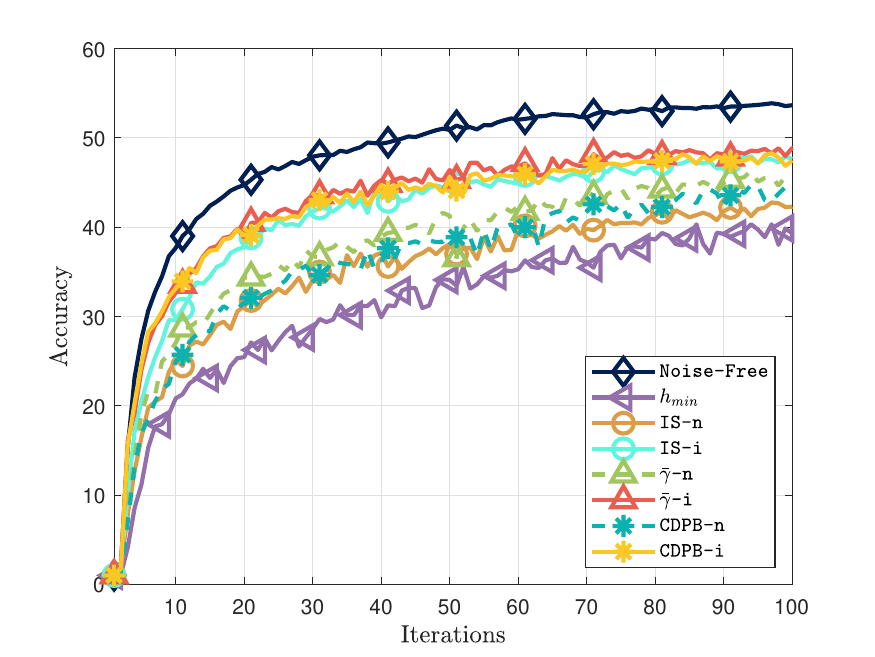}
    \caption{Test accuracy of OTA-FL with different power balancing strategies.}
    \label{Accuracy_CIFAR_100}
\end{figure}

\begin{figure}
    \centering
    \includegraphics[width=0.49\textwidth]{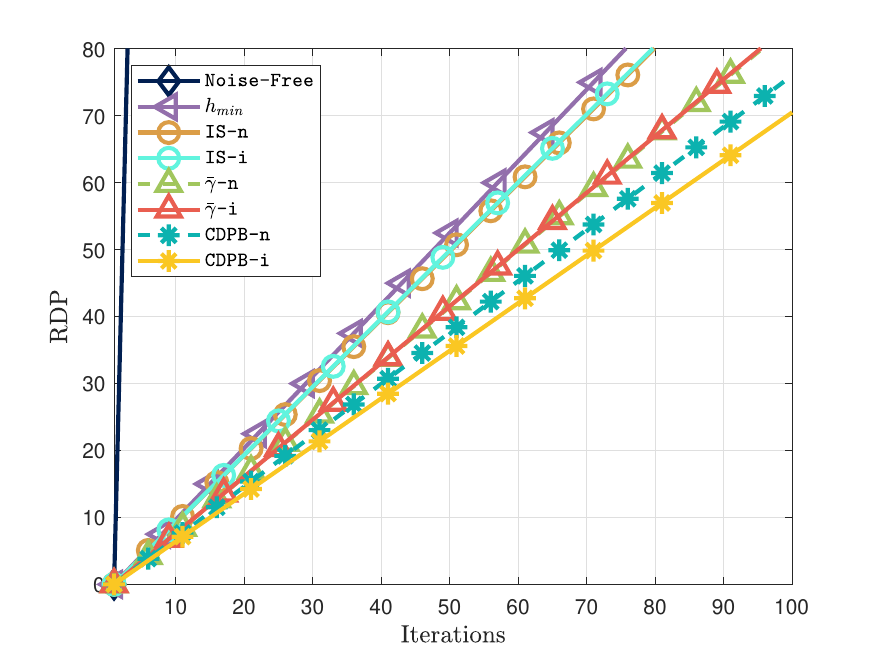}
    \caption{Achievable RDP of OTA-FL with different power balancing strategies.}
    \label{Privacy_CIFAR_100}
\end{figure}

\textbf{Test Accuracy and RDP Trade-Off}\quad
Fig. \ref{Accuracy_CIFAR_100} shows the test accuracy, while Fig. \ref{Privacy_CIFAR_100} shows the RDP with various power balancing strategies for CIFAR-100, respectively. 

Fig. \ref{Accuracy_CIFAR_100} shows that \texttt{CDPB} achieves accuracy comparable to that of the \texttt{$\bar{\gamma}$ based} approach, while converging faster than the \texttt{$h_{\text{min}}$ based} method. This is because, in the \texttt{$h_{\text{min}}$ based} strategy, all clients must transmit gradient signals, which typically requires setting $\rho$ to a lower value due to the client with the worst channel condition. This lower $\rho$ results in more noise being added, leading to reduced accuracy compared to other power balancing strategies. Meanwhile, the \texttt{IS based} method shows slightly lower accuracy than \texttt{CDPB}, as fewer clients transmit gradients. Some clients that meet the channel threshold opt out of the iteration to preserve privacy through sampling, contributing to the drop in accuracy.

Fig. \ref{Privacy_CIFAR_100} highlights that despite the increased artificial noise in the \texttt{$h_{\text{min}}$ based} strategy, it does not improve RDP compared to \texttt{CDPB} because it neglects the RDP achievable through the probability of client participation. As a result, it provides the lowest RDP guarantee among the strategies considered. Additionally, in \texttt{IS based}, since the server can access information on participating clients via CSI feedback in each iteration, it cannot fully leverage the uncertainty surrounding client participation, leading to lower RDP. For \texttt{Noise-Free}, while it ensures RDP without injecting artificial noise during training, it achieves higher accuracy than any other approach due to the absence of noise. However, its noiseless nature results in greater RDP leakage.

We also observe that \texttt{CDPB-i} outperforms \texttt{CDPB-n} in both accuracy and RDP. This is because RDP is influenced not only by artificial noise but also by the probability of client participation. Although \texttt{CDPB-n} introduces more noise than \texttt{CDPB-i}, the tighter RDP guarantees of \texttt{CDPB-i} stem from the stronger privacy protection provided by client participation probabilities, which outweigh the impact of the artificial noise in \texttt{CDPB-n}. Consequently, it's difficult to claim that \texttt{CDPB-n} offers greater privacy than \texttt{CDPB-i}. 

\textbf{Impacts of $\lambda$} \quad
\begin{table}[t!]
  \caption{$\rho$ with respect to $\lambda_2$ for $\lambda_1=1$ and $\tau=100$\\ (all values of $\lambda_2$ are in units of $10^{-5}$).}
  \label{lambda}
  \centering
  \begin{tabular}{cllll}
    \hline
    \textbf{$\lambda_2$}       & \textbf{0.5}  & \textbf{1}    & \textbf{1.5}  & \textbf{2} \\ \hline
    \textbf{\texttt{noisy}}  & 0.46 & 0.51 & 0.58 & 0.65 \\
    \textbf{\texttt{idle}}   & 0.52 & 0.60 & 0.70 & 0.79 \\
    \hline
  \end{tabular}
\end{table}
In the proposed framework, the utility function $\mathcal{G}(\rho,\tau) = \lambda_1\gamma + \lambda_2\epsilon$ balances convergence error and privacy. The hyperparameters $\lambda_1$ and $\lambda_2$ determine the relative importance assigned to convergence performance and privacy protection. Table~\ref{lambda} illustrates how the normalized power balancing parameter $\rho$ varies with $\lambda_2$ when $\lambda_1$ is fixed at 1 and $t=100$. In our framework, a lower $\rho$ corresponds to a lower channel threshold, allowing more clients to participate in gradient transmission. While increased participation typically reduces convergence error, it also heightens the risk of privacy leakage. As $\lambda_2$ increases, greater emphasis is placed on privacy in the utility function, leading to a higher $\rho$. Furthermore, we observe that $\rho$ is lower in the \texttt{noisy} compared to the \texttt{idle}. In the \texttt{noisy}, CwP transmits artificial Gaussian noise instead of remaining idle, which helps satisfy the privacy constraint although the number of clients sending gradients is reduced. This allows the system to operate with a lower channel threshold, thereby increasing client participation in gradient transmission relative to the \texttt{idle}.

\textbf{Comparison of \texttt{CDPB-n} and \texttt{CDPB-i}}\quad
The most significant difference between \texttt{CDPB-n} and \texttt{CDPB-i} lies in whether CwP send artificial noise. In Section \ref{sec:client driven power balancing}, we derived the optimal parameters $\rho_{opt}$ and $\tau_{opt}$ to achieve the best performance under privacy and acceptable convergence error constraints, using a two-stage approach. The first stage involved identifying a feasible set $\mathcal{T}$, while the second stage focused on finding the values of $\rho_{opt}$ and $\tau_{opt}$ that minimize $\mathcal{G}(\rho, \tau)$. This subsection analyzes the differences in determining $\rho_{opt}$ and $\tau_{opt}$ to compare \texttt{CDPB-n} and \texttt{CDPB-i}.

\begin{figure}[!t]
    \centering
    \subfigure[]{\includegraphics[width=0.485\linewidth]{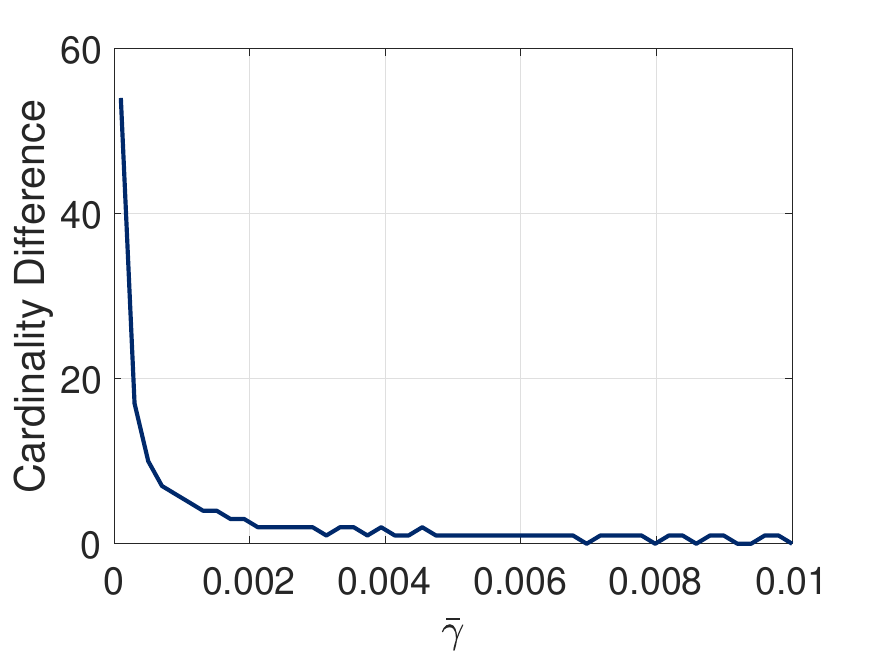}\label{fig:cardinality_disparity_gamma}}
    \subfigure[]{\includegraphics[width=0.485\linewidth]{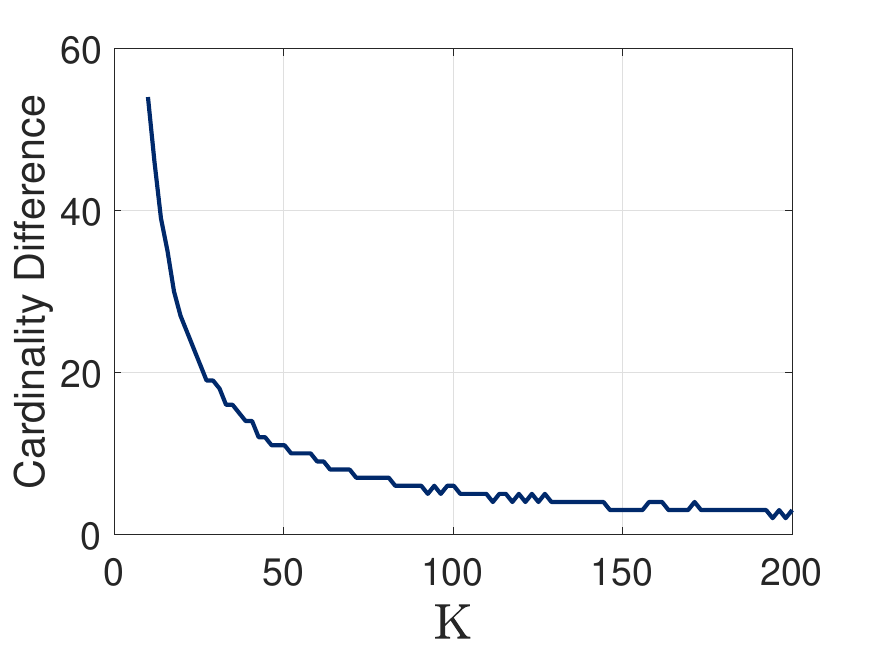}\label{fig:cardinality_disparity_K}}
    \caption{Cardinality disparity between \texttt{CDPB-n} and \texttt{CDPB-i}.}
    \label{fig:cardinality_disparity}
\end{figure}
We first address the disparity in defining the feasible set $\mathcal{T}$. The set $\mathcal{T}$ is determined by the parameters $\tau_{\gamma, \text{min}}$ and $\tau_{\epsilon, \text{max}}$, with $\tau_{\epsilon, \text{max}}$ being the same for both \texttt{CDPB-n} and \texttt{CDPB-i}. Thus, the difference in the size of $\mathcal{T}$ is solely attributed to variations in $\tau_{\gamma, \text{min}}$. As illustrated in Fig. \ref{fig:cardinality_disparity}, the disparity in $|\mathcal{T}|$ is influenced by different parameters. Figure \ref{fig:cardinality_disparity_gamma} shows that as the acceptable convergence error $\bar{\gamma}$ increases, the impact on disparity decreases due to reduced noise on the convergence error $\gamma$, narrowing the gap. Similarly, Fig. \ref{fig:cardinality_disparity_K} shows that as the total number of participating clients $K$ increases, the disparity also decreases, because increasing $K$ amplifies the effect on $\gamma$ more than the noise injection difference between the two approaches.

\begin{figure}[!t]
    \centering
    \subfigure[]{\includegraphics[width=0.485\linewidth]{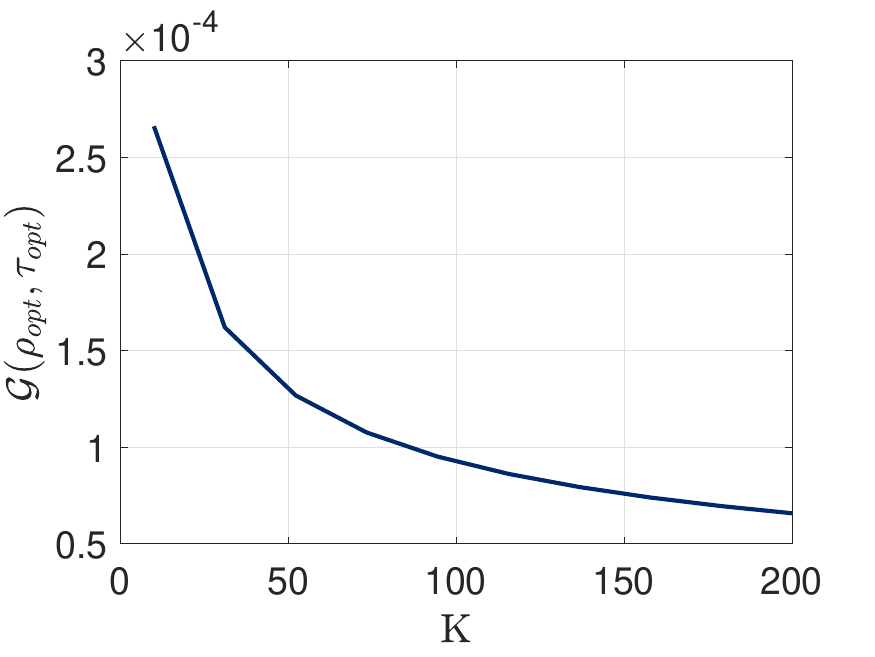}\label{fig:gain_disparity_K}}
    \subfigure[]{\includegraphics[width=0.485\linewidth]{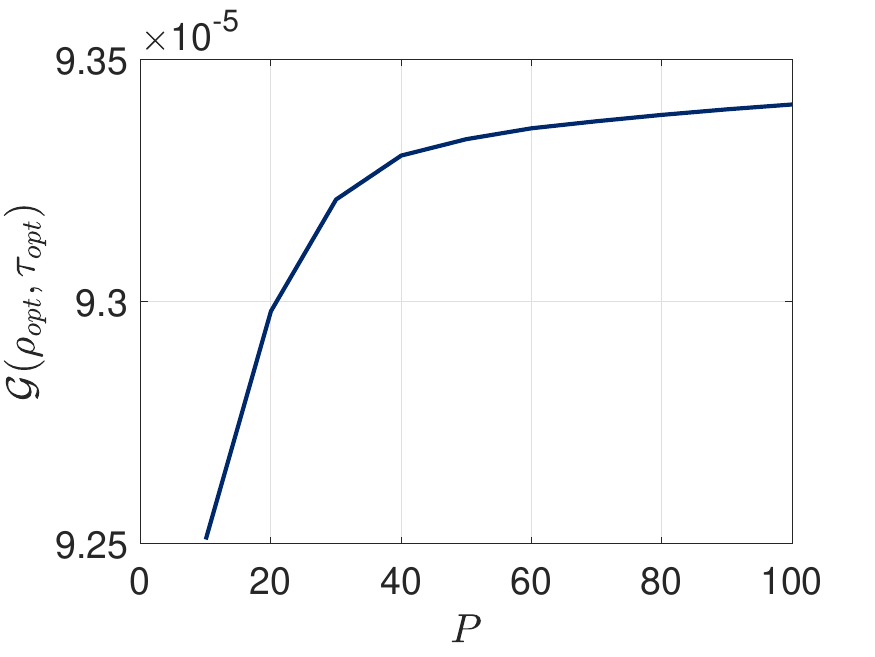}\label{fig:gain_disparity_P}}
    \caption{$\mathcal{G}(\rho_{opt},\tau_{opt})$ disparity between \texttt{CDPB-n} and \texttt{CDPB-i}.}
    \label{fig:gain_disparity}
\end{figure}
Next, we examine the disparity in utility gain $\mathcal{G}(\rho_{opt},\tau_{opt})$, shown in Fig. \ref{fig:gain_disparity}. In Fig. \ref{fig:gain_disparity_K}, the disparity decreases with an increasing number of participating clients $K$ but remains positive, indicating that \texttt{CDPB-i} consistently outperforms \texttt{CDPB-n}. A similar trend is observed with increasing available power $P$ in Fig. \ref{fig:gain_disparity_P}. This performance gap arises from differences in their privacy mechanisms. \texttt{CDPB-n} fails to fully leverage the privacy protection potential through client participation probability, resulting in consistently lower utility gain $\mathcal{G}(\rho_{opt}, \tau_{opt})$ compared to \texttt{CDPB-i}. Interestingly, the disparity in utility gain varies with changes in $K$ and $P$. As $K$ increases, the disparity decreases, but as $P$ increases, the disparity grows. This can be explained by the influence of $K$ and $P$ on the convergence error $\gamma$. Since $\sigma_{q,t}^2$ and $K^2$ are inversely proportional in the expression for $\gamma$, increasing $K$ reduces the impact of artificial noise, thereby decreasing the disparity. On the other hand, increasing $P$ leads to more noise in \texttt{CDPB-n}, widening the performance gap.

\begin{figure}[!t]
    \centering
    \subfigure[]{\includegraphics[width=0.485\linewidth]{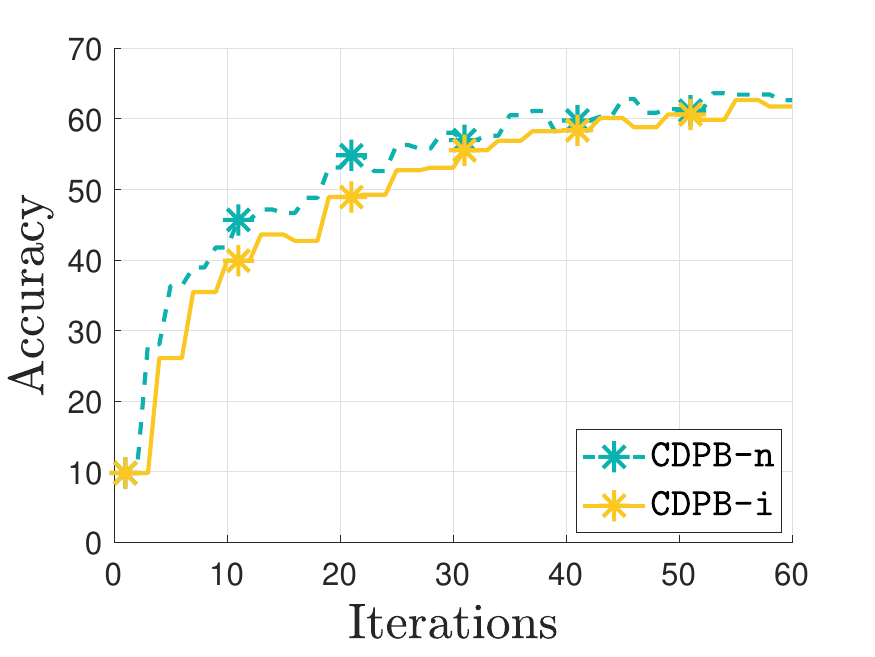}\label{fig:channel_distribution_1}}
    \subfigure[]{\includegraphics[width=0.485\linewidth]{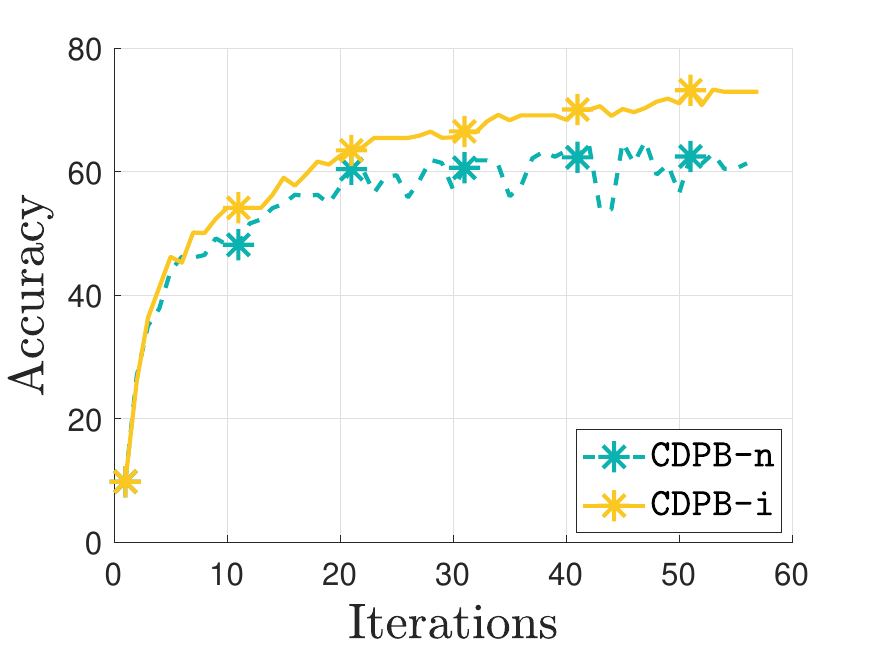}\label{fig:channel_distribution_10}}
    \caption{Test accuracy of \texttt{CDPB-n} and \texttt{CDPB-i} based OTA-FL with changes in channel distribution.}
    \label{fig:channel_distribution}
\end{figure}
The main distinction between \texttt{CDPB-n} and \texttt{CDPB-i} also extends to the complexity of calculating $\rho$. Specifically, \texttt{CDPB-i} has higher computational complexity, leading to longer times to reach a solution. Since $\rho$ needs to be recalculated whenever the channel distribution changes, \texttt{CDPB-n} may be more suitable for environments where the channel distribution fluctuates frequently. This is because \texttt{CDPB-n} computes $\rho$ more quickly than \texttt{CDPB-i}, due to a smaller feasible set $|\mathcal{T}|$ in the first stage. Fig. \ref{fig:channel_distribution} shows performance for CIFAR-10 classification across 10 clients when the channel distribution changes during training. In Fig. \ref{fig:channel_distribution_1}, where the channel distribution shifts at every iteration and $\rho$ is recalculated at each step, \texttt{CDPB-n} computes $\rho$ faster, resulting in quicker convergence compared to \texttt{CDPB-i}. Conversely, in Fig. \ref{fig:channel_distribution_10}, where $\rho$ is recalculated every 10 iterations, \texttt{CDPB-i} proves to be more suitable.
\begin{figure}[!t]
    \centering
    \includegraphics[width=0.48\textwidth]{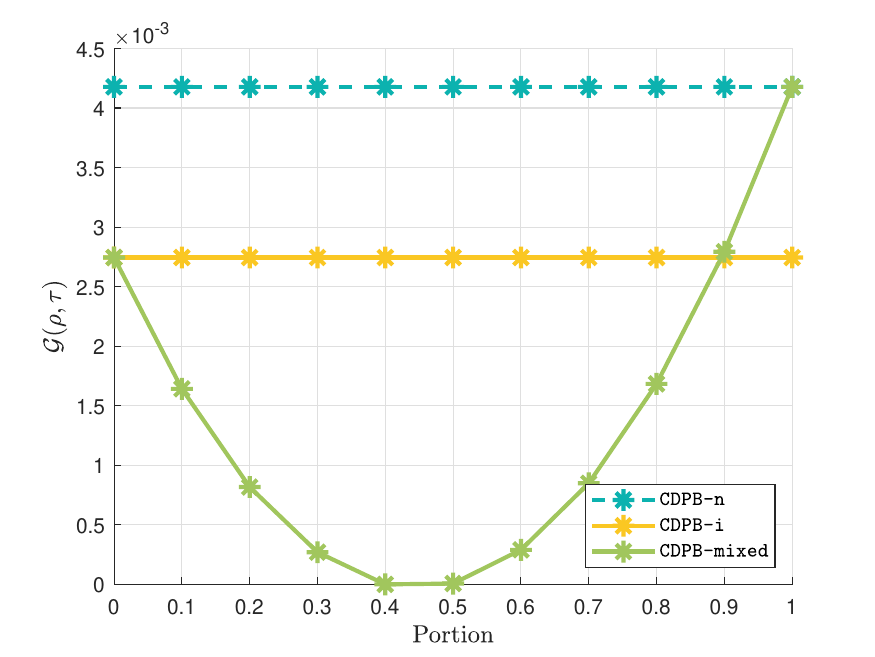}
    \caption{$\mathcal{G}(\rho, \tau)$ under \texttt{CDPB-n}, \texttt{CDPB-i} and \texttt{CDPB-mixed}.}
    \label{New_CDPB}
\end{figure}

To explore potential improvements, we revisit the observation that while \texttt{CDPB-n} introduces more noise than \texttt{CDPB-i}, \texttt{CDPB-i} offers tighter RDP guarantees. This is because the privacy protection gained from the client participation in \texttt{CDPB-i} outweighs the artificial noise injected in \texttt{CDPB-n}. However, the excessive noise in \texttt{CDPB-n} may be due to CwP using their full power $P$, as described in \eqref{eq:power_constraint}. To address this, we propose a mixed strategy, \texttt{CDPB-mixed}, combining the strengths of both \texttt{CDPB-n} and \texttt{CDPB-i}. Fig. \ref{New_CDPB} shows CwP using a coin-flipping strategy to probabilistically decide whether to transmit noise. The ``portion" represents the Bernoulli probability that CwP send noise: 0 corresponds to \texttt{CDPB-i} (no noise), while 1 corresponds to \texttt{CDPB-n} (all clients transmit noise). By adjusting this proportion, \texttt{CDPB-mixed} strikes a balance between privacy enhancement from noise and improved convergence performance. While this paper demonstrates the potential of this mixed approach, determining the optimal coin-flipping probability is an open question for future research. 

\section{Conclusion}\label{sec:conclusion}
In this paper, we proposed the \texttt{CDPB} strategy for privacy-enhanced OTA-FL, enabling clients to adjust transmission power independently based on channel distribution, eliminating the need for constant CSI feedback. We introduced two approaches for CwP: \texttt{CDPB-n}, where the CwP transmit full-power artificial Gaussian noise instead of transmitting their gradients, \texttt{CDPB-i}, where the CwP pause transmission entirely until the channel improves. We derived closed-form convergence and privacy bounds under both convex and non-convex objectives, and formulated a two-stage client-side optimization algorithm. Our results show that for the same convergence error, \texttt{CDPB-i} achieves a strictly tighter RDP guarantee than \texttt{CDPB-n}, while also delivering superior model accuracy and power efficiency by fully exploiting participation-based privacy amplification. However, \texttt{CDPB-n} is more suitable in environments with frequent channel changes due to its faster computation. We also introduced a mixed strategy, \texttt{CDPB-mixed}, where CwP probabilistically decide whether to transmit noise or remain idle. Experiments confirmed its potential to enhance learning and privacy in OTA-FL, with further exploration planned for future research.
\appendices
\section{Proof of Theorem \ref{renyi_theorem}}\label{AppendixA}
To begin with, recall that $\sigma^2$-Gaussian mechanism achieves $\left(\alpha,\frac{\alpha \Delta^2}{2\sigma^2}\right)$-RDP from Remark \ref{eq:Gaussian&RDP}. To leverage the Gaussian mechanism, we proceed by establishing an upper bound on the sensitivity. This sensitivity can be effectively bounded as $\Delta^2=\lVert \hat{g}_t-\hat{g}_t' \rVert^2=\lVert \hat{g}_t\rVert^2+\lVert \hat{g}_t'\rVert^2 \leq 2W^2.$
Therefore, with the noise injected by the clients and channel distortion $\sigma^2_{q_t}$, every client satisfies $\left(\alpha, \frac{\alpha W^2}{\sigma^2_{q_t}}\right)$-RDP at every global iteration.
In the following, we consider lemmas:
\begin{lemma}\label{rdp}
    Suppose each client participates in the OTA-FL with probability $0\leq p\leq1$ at each global iteration. For all integers $\alpha \geq 2$ and defining $E_{\alpha}=e^{\frac{\alpha W^2}{\sigma_{q_t}^2}}$, every client achieves $(\alpha,\epsilon')$-RDP with $\sigma^2_{q_t}$-Gaussian mechanism, where
\begin{align} \label{eq:epslion}
    \epsilon' = \frac{1}{\alpha-1} \log &\left( 1 + p^2 \binom{\alpha}{2} \min \left\{ 4(E_2 - 1), 2E_2 \right\} \right. \nonumber \\
    &\quad\quad\quad\quad\quad+ \left. \sum_{j=3}^{\alpha} p^j \binom{\alpha}{j} 2E_{(j-1)j} \right).
\end{align}

\end{lemma}
\begin{IEEEproof}
        The proof is in Theorem 9 of \cite{wang2019subsampled}.
\end{IEEEproof}
Next, we separate cases for the expression $\min\{4(E_2-1),2E_2\}$. Firstly, for $E_2<2$, 
\begin{align}
    \epsilon'&=\frac{1}{\alpha-1}\log\left(\sum_{j=0}^{\alpha}p^j\binom{\alpha}{j}2E_{(\alpha-1)j}+2E_2-4\right)\\
    &\overset{(a)}{=}\frac{1}{\alpha-1}\log\left(2\left(pE_{(\alpha-1)}+1\right)^\alpha+2E_2-4\right)\\
    &\overset{(b)}{\leq}\frac{1}{\alpha-1}\log\left(2\left(pE_{(\alpha-1)}+1\right)^\alpha\right),\label{eq: epsilon'}
\end{align}
where $(a)$ follows from the binomial expansion, i.e. $(x+y)^n=\sum_{k=0}^{n}\binom{n}{k}x^{n-k}y^{k}$, and $(b)$ follows $E_2-2<0$. Next for the second case, $E_2>2$, using binomial expansion,
\begin{align}
    \epsilon'=\frac{1}{\alpha-1}\log\left(2\left(pE_{(\alpha-1)}+1\right)^\alpha\right).
\end{align}
We can conclude that \eqref{eq: epsilon'} always holds.
Finally, by expanding \eqref{eq: epsilon'} to $\tau$ iterations using the composition theorem in Remark \ref{Remark1}, we can obtain Theorem \ref{renyi_theorem}.

\section{Proof of Theorem \ref{convergence_theorem}}\label{AppendixB}
In the following, we first show the relationship between $\theta_{t+1}$ and $\theta^*$ in lemma \ref{one-round_convergence} below, where $\theta_t$ shows the averaged weight throughout the participating clients described in \eqref{average-weight} and $\theta^*$ represents the optimal weight. After that, we show the relation between $f(\theta_{\tau})$ and $f(\theta^*)$ extending the lemma \ref{one-round_convergence}, which proves the theorem \ref{convergence_theorem}.

We first show how the global weight is obtained from $K$ clients. Based on the estimated gradient $\hat{g}_t$ in $\eqref{received_1}$, the server could get the estimated global weight at every $t$ iteration as it knows the previous model and initial parameter $\theta_0$. The averaged weight is given by:
\begin{align}\label{average-weight}
    \theta_{t+1}=\theta_t-\eta_t \left(\frac{1}{K_t}\sum_{k\in\mathcal{K}_t}\sum_{\ell=1}^L\nabla{f_k(\theta_{k,t,\ell})}+\frac{q_t}{K_t}\right),
\end{align}
and we define an auxiliary variable averaged weight with unbiasedness as
\begin{align}
    \zeta_{t+1}=\theta_t-\eta_t \left(\frac{1}{K_t}\sum_{k=1}^{K_t}\sum_{\ell=1}^L\nabla{f_k(\theta_{k,t,\ell})}+\frac{q_t}{K_t}\right).
\end{align}
\begin{lemma}\label{one-round_convergence}
When assumptions 1, 2, and 3 hold and the averaged weight $\theta_{t}$ is updated as \eqref{average-weight} with a learning rate $\eta_t\leq \frac{1}{6M}$, the following inequality holds:
\begin{align}
    \mathbf{E}&[\lVert\theta_{t+1}-\theta^*\rVert^2] \leq (1-\mu\eta_t)\mathbf{E}[\lVert\theta_t-\theta^*\rVert^2] + \frac{4}{K_t}\eta_t^2L^2G^2 \nonumber\\
    &L^4M^2\eta_t^4G^2+3\eta_t^2\sigma_G^2-\eta_t \mathbf{E}[f(\theta_t)-f(\theta^*)] + \eta_t^2 \frac{\sigma_{q_t}^2}{K_t^2\rho}.
\end{align}
\end{lemma}

\begin{IEEEproof}
First, we represent $\lVert\theta_{t+1}-\theta^*\rVert^2$ as $\lVert\theta_{t+1}-\zeta_{t+1}+\zeta_{t+1}-\theta^*\rVert^2=\lVert\theta_{t+1}-\zeta_{t+1}\rVert^2+\lVert\zeta_{t+1}-\theta^*\rVert^2+\langle \theta_{t+1}-\zeta_{t+1},\zeta_{t+1}-\theta^*\rangle$. Then, for the last term,  $\mathbf{E}[\langle\theta_{t+1}-\zeta_{t+1},\zeta_{t+1}-\theta^*\rangle]$ goes to zero due to the unbiasedness. Also, using the update rule,
\begin{align}\label{update rule}
\lVert\zeta_{t+1}&\!-\!\theta^*\rVert^2\!=\!\lVert\theta_t\!-\!\theta^*\rVert^2\!+\!\eta_t^2\left\lVert\frac{1}{{K_t}} \sum_{k=1}^{K_t}\sum_{\ell=1}^L\nabla{f_k(\theta_{k,t,\ell})}\!+\!q_t\right\rVert^2\nonumber\\
\quad\quad&+2\eta_t\left\langle \theta^*-\theta_t,\frac{1}{{K_t}}\sum_{k=1}^{K_t}\sum_{\ell=1}^L\nabla{f_k(\theta_{k,t,\ell})}+q_t\right\rangle.
\end{align}
For the last term on the RHS of \eqref{update rule},
\begin{align}
&2\eta_t\left\langle \theta^*-\theta_t,\frac{1}{{K_t}}\sum_{n=1}^{K_t}\sum_{\ell=1}^L\nabla{f_k(\theta_{k,t,\ell})}+q_t\right\rangle\nonumber\\
&=\frac{2\eta_t}{{K_t}}\sum_{n=1}^{K_t}\sum_{\ell=1}^L\left\langle \theta^*\!-\!\theta_t,\nabla{f_k(\theta_{k,t,\ell})}\right\rangle\!+\!2\eta_t\langle \theta^*\!-\!\theta_t,q_t\rangle,    
\end{align}
where $\langle \theta^*-\theta_t,q_t\rangle=0$. Using $\mu$-strongly convexity, 
\begin{align}\label{mu-strongly}
    2\eta_t&\frac{1}{{K_t}}\sum_{k=1}^{K_t}  \sum_{\ell=1}^L \mathbf{E}\left[\left\langle \theta^*-\theta_t,\nabla{f_k(\theta_{k,t,\ell})}\right\rangle\right]\nonumber\\
    &\leq 2\eta_t  \mathbf{E}[f(\theta^*)-f(\theta_t)]
    -\mu\eta_t \mathbf{E}[\lVert\theta_t-\theta^*\rVert^2].
\end{align}
For the second term on the RHS of \eqref{update rule}, we also have 
\begin{align}\label{noise-bound}
\eta_t^2&\mathbf{E}\left[\left\lVert \frac{1}{{K_t}}\sum_{k=1}^{K_t}\sum_{\ell=1}^L\nabla{f_k(\theta_{k,t,\ell})}+q_t\right\rVert^2\right]\nonumber\\
&= \eta_t^2\mathbf{E}\left[\left\lVert \frac{1}{{K_t}}\sum_{k=1}^{K_t}\sum_{\ell=1}^L\nabla{f_k(\theta_{k,t,\ell})}\right\rVert^2\right]+ \eta_t^2\frac{\sigma_{q_t}^2}{K_t^2\rho}\nonumber\\
&+ 2\eta_t^2\mathbf{E}\left[\left\langle \frac{1}{{K_t}}\sum_{k=1}^{K_t}\sum_{\ell=1}^L\nabla{f_k(\theta_{k,t,\ell})},q_t\right\rangle\right], 
\end{align}
where $\mathbf{E}[\langle \frac{1}{{K_t}}\sum_{k=1}^{K_t}\sum_{\ell=1}^L\nabla{f_k(\theta_{k,t,\ell})},q_t\rangle]=0$. For the first term of \eqref{noise-bound}, by adding and subtracting both $\nabla{f_k(\theta_{k,t})}$ and $\nabla{f(\theta_{t})}$ inside the norm, it can be expressed as:
\begin{align}
    &\eta_t^2\mathbf{E}\left[\left\lVert \frac{1}{{K_t}}\sum_{k=1}^{K_t}\sum_{\ell=1}^L\nabla{f_k(\theta_{k,t,\ell})}\right\rVert^2\right]\nonumber\\
    &= \eta_t^2\mathbf{E}\Bigg[\Bigg\|\frac{1}{{K_t}}\sum_{k=1}^{K_t}\sum_{\ell=1}^L\nabla f_k(\theta_{k,t,\ell})-\nabla f_k(\theta_{k,t})\nonumber\\
    &\qquad\;\;\;\;\;+\nabla f_k(\theta_{k,t})-\nabla f(\theta_t)+\nabla f(\theta_t)\Bigg\|^2\Bigg]\\
    &\leq \frac{3\eta_t^2L}{{K_t}}\sum_{k=1}^{K_t}\sum_{\ell=1}^L\mathbf{E}\left[\left\|\nabla f_k(\theta_{k,t,\ell})-\nabla f_k(\theta_{k,t})\right\|^2\right]\nonumber\\
    &\quad+\frac{3\eta_t^2}{{K_t}}\sum_{k=1}^{K_t}\mathbf{E}\left[\left\|\nabla f_k(\theta_{k,t})-\nabla f(\theta_t)\right\|^2\right]\nonumber\\
    &\quad+3\eta_t^2\mathbf{E}\left[\left\|\nabla f(\theta_t)\right\|^2\right]\\
    &\leq \frac{3\eta_t^2LM^2}{{K_t}}\sum_{k=1}^{K_t}\sum_{\ell=1}^L\|\theta_{k,t,\ell}-\theta_{k,t}\|^2+3\eta_t^2\sigma_G^2\nonumber\\
    &\quad+6M\eta_t^2\mathbf{E}[f(\theta_t)-f(\theta^*)],\label{leq: gradient}
\end{align}
where the first inequality is using a simple property, $\|a+b+c\|^2\leq 3\|a\|^2+3\|b\|^2+3\|c\|^2$ and Jensen's inequality. The second inequality is due to Assumption 4 and $M$-smooth. For the first term in \eqref{leq: gradient}, 
\begin{align}
    &\frac{1}{K_t}\sum_{k=1}^{K_t}\sum_{\ell=1}^L\|\theta_{k,t,\ell}-\theta_{k,t}\|^2\nonumber\\
    &=\frac{1}{{K_t}}\sum_{k=1}^{K_t}\sum_{\ell=1}^L\mathbf{E}\left[\left\|\sum_{i=1}^{\ell}\eta_t\nabla f_k(\theta_{k,t,i})\right\|^2\right]\\
    &\leq \eta_t^2\sum_{\ell=1}^L\ell^2G^2\\
    &=\frac{L(L+1)(2L+1)}{6}\eta_t^2G^2\\
    &\leq \frac{L^3}{3}\eta_t^2G^2,
\end{align}
where the first inequality is from Assumption 3 and Jensen's inequality.

Combining the above results and using 
the expected difference between $\theta_t$ and $\zeta_t$ is bounded by $\mathbf{E}[\lVert\theta_t-\zeta_t\rVert^2]\leq\frac{4}{K_t}\eta_t^2L^2G^2$ following the lemma in Appendix B.4 of \cite{li2019convergence}, we obtain

\begin{align}
    &\mathbf{E}[\lVert\theta_{t+1}-\theta^*\rVert^2] \leq (1-\mu\eta_t)\mathbf{E}[\lVert\theta_t-\theta^*\rVert^2] + \frac{4}{K_t}\eta_t^2L^2G^2 \nonumber\\
    &L^4M^2\eta_t^4G^2+3\eta_t^2\sigma_G^2-\eta_t \mathbf{E}[f(\theta_t)-f(\theta^*)] \!+\! \eta_t^2 \frac{\sigma_{q_t}^2}{K_t^2\rho}.
\end{align}

\end{IEEEproof}
\begin{lemma}\label{lemma3.4}
\textit{(Lemma 3.4 in \cite{sery2021over})} Let $\{a_t\}_{t\geq0}$ and $\{e_t\}_{t\geq0}$ be positive sequences satisfying
\begin{align}
    a_{t+1}\leq(1-\mu\eta_t)a_t-\eta_te_tA+\eta_t^2B,
\end{align}
for $\eta_t=\frac{4}{\mu(a+t)}$ with $A\geq0$ and $B, C\geq0$. For positive integer $\tau$, it holds
\begin{align}
    \frac{A}{S_\tau}\sum_{t=0}^{\tau-1}\beta_{t}e_{t}\leq\frac{\mu a^3}{4S_\tau}a_0+\frac{2\tau(\tau+2a)}{\mu S_\tau}B
\end{align}
for $\beta_t=(a+t)^2$ and $S_\tau=\sum_{t=0}^{\tau}\beta_{t}=\frac{\tau}{6}(2\tau^2+6a\tau-3\tau+6a^2-6a+1)\geq\frac{1}{3}\tau^3$.
\end{lemma}

We finally prove Theorem 2 by substituting the terms of Lemma \ref{one-round_convergence} into Lemma \ref{lemma3.4}. Setting $A=1, B=\frac{4}{K_t}M^2G^2+\frac{\sigma_{q_t}^2}{K_t^2\rho},e_t=\mathbf{E}[f(\theta_t)-f(\theta^*)]$, 

\begin{align}
    \mathbf{E}&[f(\Theta_\tau)]-f(\theta^*)\leq\frac{\mu a^3}{4S_\tau}\mathbf{E}[\lVert\theta_0-\theta^*\rVert^2]\nonumber\\
    &+ \frac{2\tau(\tau+a)}{\mu S_\tau}\left(\frac{4}{K_t}L^2G^2+\frac{\sigma_{q_t}^2}{K_t^2\rho}+L^4M^2\eta_t^2G^2+3\sigma_G^2\right)
\end{align}

where $\Theta_\tau=\frac{1}{S_\tau}\sum_{t=0}^{\tau-1}\beta_t\theta_{t}, $ for $\beta_t=(a+t)^2, S_\tau=\sum_{t=0}^{\tau-1}\beta_{t}\geq\frac{1}{3}\tau^3.$
\section{Proof of Theorem \ref{convergence_theorem_nonconvex}}\label{AppendixC}
Similar to the strong convexity case, we begin by introducing an auxiliary variable to account for the impact of partial participation, representing the unbiased averaged weight. Specifically, we define
\begin{align}
    \zeta_{t+1}=\theta_t-\eta_t \left(\frac{1}{K_t}\sum_{k=1}^{K_t}\sum_{\ell=1}^L\nabla{f_k(\theta_{k,t,\ell})}+\frac{q_t}{K_t}\right).
\end{align}
Under the smoothness assumption, the expected change in the objective function between two consecutive iterates can be written as
\begin{align}\label{smooth}
    \mathbf{E}[f(\theta_{t+1})-f(\theta_{t})]&\leq\underbrace{\mathbf{E}[\langle \nabla f(\theta_t),\theta_{t+1}-\theta_t \rangle]}_{(a)}\nonumber\\
    &+\frac{M}{2}\underbrace{\mathbf{E}[\|\theta_{t+1}-\theta_t\|^2]}_{(b)}.
\end{align}
An upper bound on $(a)$ in \eqref{smooth} can be obtained as  
\begin{align}
    (a)&= -\eta_t \sum_{\ell=1}^{L}\mathbf{E}\left[\left\langle \nabla f(\theta_t), \frac{1}{{K_t}}\sum_{k=1}^{K_t} \nabla f_k(\theta_{k,t,\ell})\right \rangle\right]\nonumber\\
    &= -\frac{\eta_t}{2}\underbrace{\sum_{\ell=1}^L \mathbf{E}\left[\left\|\frac{1}{{K_t}}\sum_{k=1}^{K_t}\nabla f_k(\theta_{k,t,\ell})
    \right\|^2\right]}_{(a.1)}\nonumber\\
    &+\frac{\eta_t}{2}\underbrace{\sum_{\ell=1}^L\mathbf{E}\left[\left\|\nabla f(\theta_t) - \frac{1}{{K_t}}\sum_{k=1}^{K_t} \nabla f_k(\theta_{k,t,\ell})\right\|^2\right]}_{(a.2)},
\end{align}
where the first equality comes from unbiasedness, the second equality is due to $-a^Tb=\frac{1}{2}(-\|a\|^2-\|b\|^2+\|a-b\|^2)$. Because the terms inside the summation are non-negative, we get a lower bound $(a.1)$ by choosing only $\ell=1$ from the summation of positive norm values, 
\begin{align}
    (a.1)&\geq \mathbf{E}\left[\left\|\frac{1}{{K_t}}\sum_{k=1}^{K_t} \nabla f_k(\theta_{k,t,1}))\right\|^2\right]\\
    &= \mathbf{E}\left[\left\| \nabla f(\theta_t)\right\|^2\right].
\end{align}
For $(a.2)$, using smoothness and Jensen's inequality, 
\begin{align}
    (a.2)&\leq M^2\frac{1}{{K_t}}\sum_{k=1}^{K_t} \sum_{\ell=1}^L\mathbf{E}\left[\left\|\theta_{k,t,1}-\theta_{k,t,\ell}\right\|^2\right]\nonumber\\
    &=M^2\frac{1}{{K_t}}\sum_{k=1}^{K_t}\sum_{\ell=1}^L\mathbf{E}\left[\left\|\sum_{i=1}^{\ell}\eta_t\nabla f_k(\theta_{k,t,i})\right\|^2\right]\nonumber\\
    &\leq M^2\eta_t^2\sum_{\ell=1}^L\ell^2G^2\nonumber\\
    &=\frac{L(L+1)(2L+1)}{6}M^2\eta_t^2G^2\nonumber\\
    &\leq \frac{L^3}{3}M^2\eta_t^2G^2,
\end{align}
where the second inequality comes from Jensen's inequality and Assumption 3.

We next turn our attention to the second term $(b)$ in \eqref{smooth}. By decomposing the squared norm as $\|\theta_{t+1}-\theta_t\|^2$ as $\|\theta_{t+1}-\zeta_{t+1}\|^2+\|\zeta_{t+1}-\theta_{t}\|^2+2\langle\theta_{t+1}-\zeta_{t+1},\zeta_{t+1}-\theta_t\rangle$. Then, for the last term, $\mathbf{E}[\langle\theta_{t+1}-\zeta_{t+1},\zeta_{t+1}-\theta_t\rangle]$ goes to zero due to the unbiasedness. Similar to the strongly convex setting, we use the expected difference between $\theta_t$ and $\zeta_t$ is bounded by $\mathbf{E}[\lVert\theta_t-\zeta_t\rVert^2]\leq\frac{4}{K_t}\eta_t^2L^2G^2$ following the lemma in Appendix B.4 of [25]]. Next, for the $\|\zeta_{t+1}-\theta_{t}\|^2$, we can derive an upper bound using
\begin{align}
    \lVert\zeta_{t+1}\!-\!\theta_t\rVert^2\!&=\!\eta_t^2\left\lVert\frac{1}{{K_t}} \sum_{k=1}^{K_t}\sum_{\ell=1}^L\nabla{f_k(\theta_{k,t,\ell})}\!+\!q_t\right\rVert^2\nonumber\\
    &= \eta_t^2\underbrace{\mathbf{E}\left[\left\lVert \frac{1}{{K_t}}\sum_{k=1}^{K_t}\sum_{\ell=1}^L\nabla{f_k(\theta_{k,t,\ell})}\right\rVert^2\right]}_{(b.1)}+ \eta_t^2\frac{\sigma_{q_t}^2}{K_t^2\rho}\nonumber\\
    &+ 2\eta_t^2\mathbf{E}\left[\left\langle \frac{1}{k}\sum_{k=1}^{K_t}\sum_{\ell=1}^L\nabla{f_k(\theta_{k,t,\ell})},q_t\right\rangle\right], 
\end{align}
where $\mathbf{E}[\langle \frac{1}{{K_t}}\sum_{k=1}^{K_t}\sum_{\ell=1}^L\nabla{f_k(\theta_{k,t,\ell})},q_t\rangle]=0$ due to unbiasedness. 

Also, by adding and subtracting both $\nabla{f_k(\theta_{k,t})}$ and $\nabla{f(\theta_{t})}$ inside the norm, $(b.1)$ can be expressed as:
\begin{align}
    (b.1)&= \mathbf{E}\Bigg[\Bigg\|\frac{1}{{K_t}}\sum_{k=1}^{K_t}\sum_{\ell=1}^L\nabla f_k(\theta_{k,t,\ell})-\nabla f_k(\theta_{k,t})\nonumber\\
    &\qquad\;\;\;\;\;+\nabla f_k(\theta_{k,t})-\nabla f(\theta_t)+\nabla f(\theta_t)\Bigg\|^2\Bigg]\\
    &\leq \frac{3L}{{K_t}}\sum_{k=1}^{K_t}\sum_{\ell=1}^L\mathbf{E}\left[\left\|\nabla f_k(\theta_{k,t,\ell})-\nabla f_k(\theta_{k,t})\right\|^2\right]\nonumber\\
    &\quad+\frac{3}{{K_t}}\sum_{k=1}^{K_t}\mathbf{E}\left[\left\|\nabla f_k(\theta_{k,t})-\nabla f(\theta_t)\right\|^2\right]\nonumber\\
    &\quad+\frac{3}{{K_t}}\sum_{k=1}^{K_t}\mathbf{E}\left[\left\|\nabla f(\theta_t)\right\|^2\right]\\
    &\leq L^4M^2\eta_t^2G^2+3\sigma_G^2+3G^2,
\end{align}
where the first inequality is using a simple property, $\|a+b+c\|^2\leq 3\|a\|^2+3\|b\|^2+3\|c\|^2$ and Jensen's inequality. The second inequality is due to Assumptions 3, 4, and using the same steps applied in $(a.2)$.

Substituting the above inequalities, which contain various bounds to \eqref{smooth}, we can derive
\begin{align}
    \mathbf{E}[&f(\theta_{t+1})-f(\theta_{t})]\leq -\frac{\eta_t}{2}\mathbf{E}\left[\left\|\nabla f(\theta_t)\right\|^2\right]+\frac{L^3}{6}M^2\eta_t^3G^2\nonumber\\
    &+ \frac{\eta_t^2M}{2}\left(\frac{4L^2G^2}{K_t}\!+\!\frac{\sigma_{q_t}^2}{K_t^2\rho}\!+\!L^4M^2\eta_t^2G^2+3\sigma_G^2+3G^2\right).
\end{align}
Averaging the above inequality over iteration from $0$ to $\tau-1$,

\begin{align}
    &\frac{1}{\tau}\sum_{t=0}^{\tau-1}\mathbf{E}\left[\|\nabla f(\theta_t)\|^2\right] \leq \sum_{t=0}^{\tau-1}\eta_t^2\frac{L^3}{3\tau}M^2G^2  + \frac{2}{\tau}[f(\theta_0)-\theta^*] \nonumber\\
    &\!+\!\frac{M}{\tau}\sum_{t=0}^{\tau-1}\eta_t\left(\frac{4L^2G^2}{K_t} \!+\! \frac{\sigma_{q_t}^2}{K_t^2\rho} \!+\! L^4M^2\eta_t^2G^2+3\sigma_G^2+3G^2\right),
\end{align}
which completes the proof.
\section{Proof of Lemma \ref{lemma:1}}\label{AppendixD}
The proof begins by simplifying $\mathcal{P}_2$. According to Theorem \ref{convergence_theorem}, $\gamma$ decreases at a rate of $\tau^{-1}$. To better understand this behavior, we approximate $\gamma$ by focusing on the dominant factors contributing to its convergence error. Given that $S_{\tau}^{-1}$ scales approximately as $\tau^{-3}$, the first term of $\gamma$ diminishes quickly compared to the second term as $\tau$ increases. Consequently, the second term becomes predominant in the long run, allowing us to approximate $\gamma$ as $\gamma \approx \frac{6}{\tau} \left(\frac{4M^2G^2}{K_t} + \frac{\sigma^2_{q_t}}{K_t^2 \rho}\right)$. Also in the Theorem \ref{convergence_theorem_nonconvex}, we observe that the same term, $\frac{4M^2G^2}{K_t} + \frac{\sigma^2_{q_t}}{K_t^2 \rho}$, again playing the dominant role. 

In the case of the solution $\tau_{\gamma,\text{min}}$, the convergence error reaches its maximum acceptable convergence error. Therefore, the first constraint in $\mathcal{P}_2$ can be set as $\gamma=\bar{\gamma}$. After substituting this constraint into the objective function, $\mathcal{P}_2$ can be reformulated as an alternative Problem $\mathcal{P}_5$:
\begin{subequations}\label{OP:P5}
\begin{align}
\mathcal{P}_5:  &\min_{\rho}  \frac{1}{\bar{\gamma}} \left(\frac{4M^2G^2}{K_t} + \frac{\sigma^2_{q_t}}{K_t^2 \rho}\right)  \\
\text{s.t.}& \ \  \sigma_{q_t}^2 \leq 2KP+\sigma_{z_t}^2,\\
& \ \  0 \leq \rho \leq \frac{P}{W^2}.
\end{align}
\end{subequations}
To solve $\mathcal{P}_4$, we first express the expected number of participating clients, $K_t$, as a function of the power balancing parameter $\rho$. Specifically, $K_t$ is the sum of $K$ Bernoulli random variables, where the success probability $p$ represents the likelihood that a client's channel gain exceeds the threshold $h_{th}$, i.e., $K_t=\sum_{k=1}^K Pr(h_k>h_{th})$. Since clients are aware of their channel distribution, $p$ is determined by the threshold defined in \eqref{eq:rho}, $h_{th}=\frac{\rho W^2}{P}$. Assuming the channel gain follows a Rayleigh fading distribution with variance $\sigma_k^2$, 

\begin{align}
    K_t&=\sum_{k=1}^K \int_{\frac{\rho W^2}{P}}^\infty \frac{h}{\sigma_k^2}e^{-h/{2\sigma_k^2}} dh\\
    &= \sum_{k=1}^Ke^{\frac{-\rho W^2}{2P\sigma_k^2}}:=K_t(\rho).\label{eq:rho_K}
\end{align}
Next, we closely look into $\sigma_{q_t}^2$. The primary objective of this paper is to focus on power balancing allocated to the client for gradients and artificial noise. Consequently, we will proceed by disregarding the influence of channel noise from this point onward, denoted as $\sigma_{z_t}^2$, but concentrate on artificial noise. Therefore, utilizing the definition in \eqref{eq:noisy_var} and \eqref{eq:idle_var}, we can express $\sigma_{q_t}^2$ as follows.

For \texttt{CDPB-n}:
    \begin{align}
        \sigma_{q_t}^2&=d\sum_{k\in\mathcal{K}_t} \rho \sigma_{r_{k,t}}^2+d\sum_{k\not\in\mathcal{K}_t} h_{k,t}\sigma_{r_{k,t}}^2\\
        &\overset{(a)}{=}\sum_{k\in\mathcal{K}_t}h_{k,t}\left[P\!-\!\frac{\rho W^2}{h_{k,t}}\right]+\sum_{k\not\in\mathcal{K}_t}h_{k,t}P\\
        &\overset{(b)}{=}\sum_{k=1}^K2P\sigma_k^2\!-\!\rho W^2 e^{\frac{-\rho W^2}{2P\sigma_k^2}}:=P_{n}(\rho),\label{eq:rho_P_rho_K}
    \end{align}
    where $(a)$ follows from the definition \eqref{eq:precoding} and \eqref{eq:power_constraint}, $(b)$ follows from the definition of Rayleigh distribution with scale parameter $\sigma_k^2$.

For \texttt{CDPB-i}:
    \begin{align}\label{eq:P_R(rho)}
        \sigma_{q_t}^2\!&=\!d \sum_{k\in\mathcal{K}_t} \rho\sigma_{r_{k,t}}^2\\        &=\sum_{k\in\mathcal{K}_t}h_{k,t}\left[P-\frac{\rho W^2}{h_{k,t}}\right]\\
        &=\sum_{k=1}^K \mathbf{E}\left[h_kP\!-\!\rho W^2|h_k\geq \frac{\rho W^2}{P}\right]\Pr\left(h_k\geq\frac{\rho W^2}{P}\right)\\
        &\overset{(c)}{=} \sum_{k=1}^K 2P\sigma_k^2e^{\frac{-\rho W^2}{2P\sigma_k^2}}:=P_{i}(\rho),
    \end{align}
    where $(c)$ follows from the definition of Rayleigh distribution with scale parameter $\sigma_k^2$.

Next, we show the optimization problem $\tau_{\text{min}}$ is convex with respect to $\rho$. The second derivative of the objective function is non-negative, which confirms convexity. 

Furthermore, the $\sigma_{q,t}^2$ for \texttt{CDPB-n} and \texttt{CDPB-i} are both convex to $\rho$ in $0\leq\rho\leq\frac{P}{W^2}$, it implies the objective function is convex. As the optimization function is strictly convex, we can solve the problem simply by applying the Lagrange multiplier method and Karush-Kuhn-Tucker (KKT) conditions. For \texttt{CDPB-n}, the partial Lagrange function is defined as
\begin{align}
    \mathcal{L}(\rho,\psi_1,\psi_2,\psi_3)&=\frac{P_{n}(\rho)}{\bar{\gamma}(K_t(\rho))^2\rho}+\frac{4 M^2G^2}{\bar{\gamma}K_t(\rho)}-\psi_1\rho\nonumber\\
    &+\psi_2\left(\rho-P/W^2\right)-\psi_3 K_t(\rho)\rho W^2,
\end{align}
where $\psi_1\geq0, \psi_2\geq0, \psi_3\geq0$ are the Lagrange multipliers with regard to gradient power constraints and a noise power constraint. For ease of analysis, we assume $\sigma_k^2$ is all the same for clients. Then, applying the KKT conditions, we have
\begin{align}
    \rho = \frac{P\sigma^2(\sqrt{4a_n+9}-1)}{W^2(a_n+2)},
\end{align}
where $a_n=4GM^2-W^2$.

For \texttt{CDPB-i}, we can solve the problem by simply substituting $\sigma_{q_t}^2$ with \eqref{eq:P_R(rho)}.

\section{Proof of Lemma \ref{lemma:2}}\label{AppendixE}
Apparently, the solution $\tau_{\epsilon, \text{max}}$ reaches its maximum privacy requirement since $\tau$ is linearly proportional to $\epsilon$. Therefore, the first constraint in $\mathcal{P}_3$ can be set as $\epsilon=\bar{\epsilon}$. Next, we prove the $\rho$ for $\tau_{\epsilon, \text{max}}$ is always $\rho=\frac{P}{W^2}$ for both \texttt{CDPB-n} and \texttt{CDPB-i}. Given the expression $p = e^{\frac{- \rho W^2}{ 2P\sigma^2}}$ and $\sigma_{q_t}^2 =2KP\sigma^2\!+\!\sigma_{z_t}^2\!-\! e^{\frac{-\rho W^2}{2P\sigma^2}}K\rho W^2$ for \texttt{CDPB-n} and $2KP\sigma^2e^{\frac{-\rho W^2}{2P\sigma^2}}+\sigma_{z_t}^2$ for \texttt{CDPB-i}, the problem is convex in terms of $\rho$   and is monotonically increasing from $\rho=0$ to $\rho=P/W^2$. Hence, the $\tau_{\text{max}}$ is attained at the maximum $\rho$, $\rho=\frac{P}{W^2}$ for both \texttt{CDPB-n} and \texttt{CDPB-i}.

%
%
%

\ifCLASSOPTIONcaptionsoff
  \newpage
\fi

\bibliographystyle{IEEEtran}
\bibliography{IEEEabrv,references}

\begin{IEEEbiography}
	[{\includegraphics[width=1in,height=1.25in,clip,keepaspectratio]{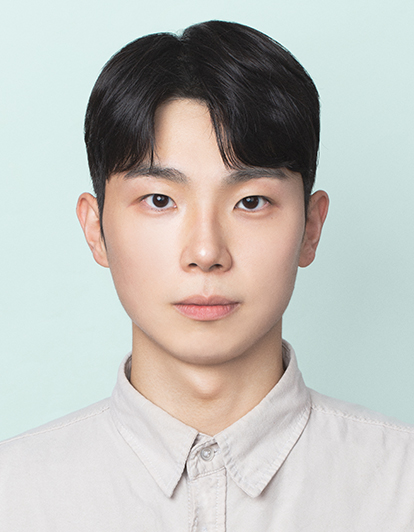}}]{Bumjun Kim}(S'20) received the B.S. degree from the Department of Electronic Engineering, Jeonbuk National University, Jeonju, South Korea, in 2020. He is currently pursuing the Ph.D. degree with the Department of Electrical and Computer Engineering, Seoul National University, Seoul. His research interests include wireless communications, distributed learning, and semantic communication.
\end{IEEEbiography}

\begin{IEEEbiography}
	[{\includegraphics[width=1in,height=1.25in,clip,keepaspectratio]{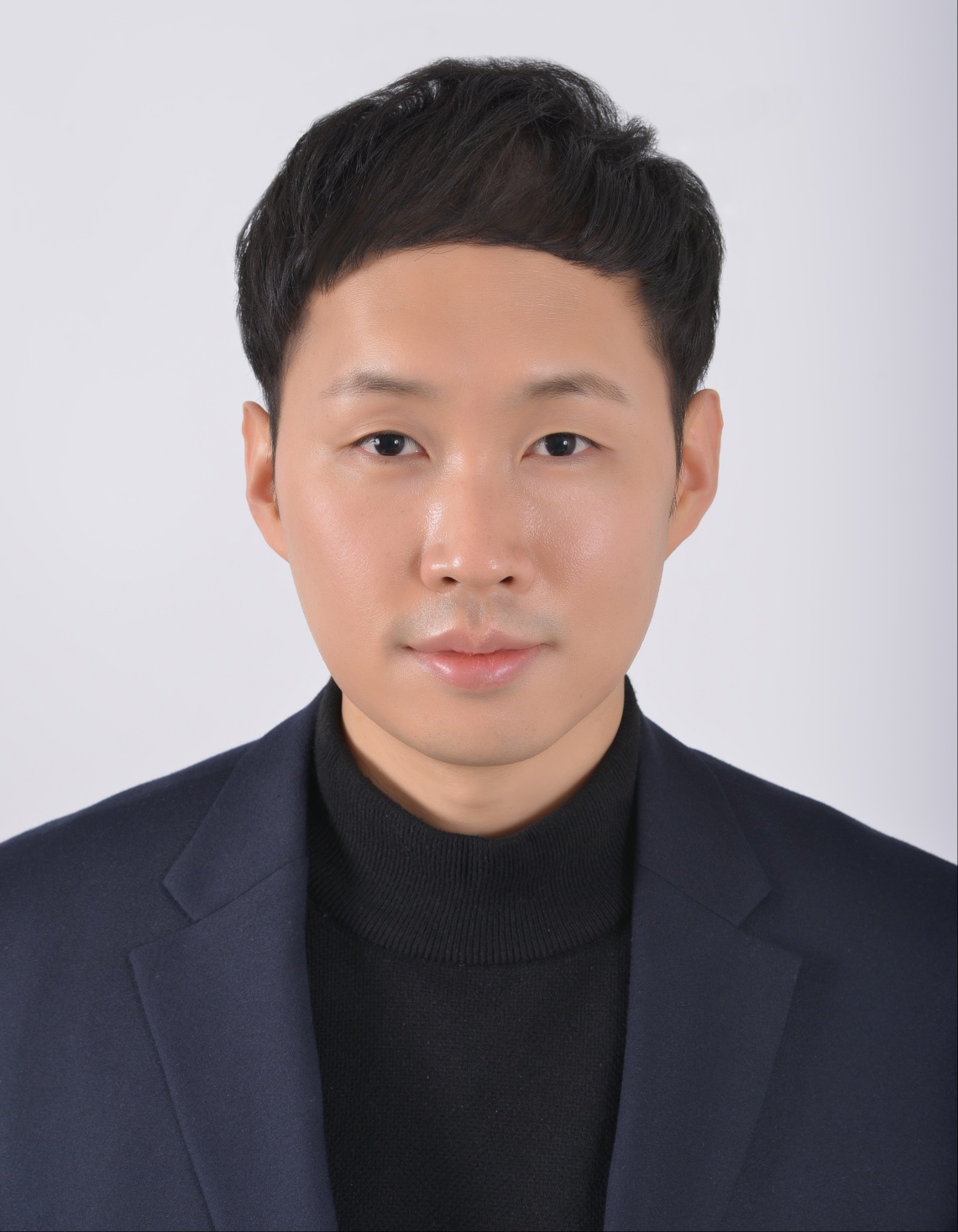}}]{Hyowoon Seo} (S’12–M’20) received the B.S., M.S., and Ph.D. degrees in electrical engineering from the Korea Advanced Institute of Science and Technology (KAIST), Daejeon, Korea, in 2012, 2014, and 2020, respectively. From 2020 to 2021, he was a postdoctoral researcher at Seoul National University and later at the Centre for Wireless Communications (CWC), University of Oulu, Finland. He then served as an assistant professor in the Department of Electronics and Communications Engineering at Kwangwoon University, Seoul, Korea. In 2025, he joined the faculty at Sungkyunkwan University, Suwon, Korea, where he is currently an assistant professor in the College of Information and Communication Engineering. Since 2025, he has also been serving as an Editor for the IEEE Transactions on Wireless Communications. His research interests include wireless communication, semantic communication, the Internet of Things (IoT), physical-layer security and privacy, vehicle-to-everything (V2X) communications, and deep and distributed learning.
\end{IEEEbiography}
\begin{IEEEbiography}
	[{\includegraphics[width=1in,height=1.25in,clip,keepaspectratio]{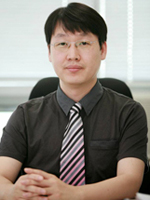}}]{Wan Choi}(S’03–M’06–SM’12-F’20) is Professor of Department of Electrical and Computer Engineering, Seoul National University, Seoul, Korea. He is Director of Institute of New Media and Communications (INMC), Seoul National University, from Feb. 2025. From Feb. 2007 to Feb. 2020, he was Professor of School of Electrical Engineering, Korea Advanced Institute of Science and Technology (KAIST), Daejeon, Korea. He received the B.Sc. and M.Sc. degrees from the School of Electrical Engineering and Computer Science (EECS), Seoul National University (SNU), Seoul, Korea, in 1996 and 1998, respectively, and the Ph.D. degree in the Department of Electrical and Computer Engineering at the University of Texas at Austin in 2006. From 1998 to 2003, he was a Senior Member of the Technical Staff of the R\&D Division of KT, Korea, where he researched 3G CDMA systems.
	
	He is the recipient of IEEE Vehicular Technology Society Jack Neubauer Memorial Award (Best System Paper Award) in 2002. He also received the IEEE Vehicular Technology Society Dan Noble Fellowship Award in 2006, the IEEE Communication Society Asia Pacific Young Researcher Award in 2007, Haedong Young Scholar Award from KICS in 2012, and the Irwin-Jacobs Award from Qualcomm and KICS in 2015. While at the University of Texas at Austin, he was the recipient of William S. Livingston Graduate Fellowship and Information and Telecommunication Fellowship from Ministry of Information and Communication (MIC), Korea. He is an Area Editor for the IEEE Transactions on Wireless Communications from Aug. 2022 and an Editor for the IEEE Transactions on Vehicular Technology from Apr. 2011. He is the co-Editor-in-Chief of IEEE/KICS Journal of Communications and Networks from Dec. 2024. He served as the Executive Editor Chair for the IEEE Transactions on Wireless Communications (2019- 2021) and Executive Editor (2014-2019). He was also an Editor for the IEEE Transactions on Wireless Communications (2009-2014), for the IEEE Wireless Communications Letter (2012-2017), and as Guest Editor for the IEEE Journal on Selected Areas in Communications.
\end{IEEEbiography}

\end{document}